\documentclass[pra,aps,twocolumn,groupedaddress]{revtex4}
\usepackage{amssymb,amsmath,amsfonts,epsfig,graphicx,latexsym,bm,color}

\newcommand{\bq}{\textbf{q}}

\begin{document}
\title{Merging and alignment of Dirac points in a shaken honeycomb optical lattice}
\author{Selma Koghee$^{1,2}$, Lih-King Lim$^{2}$, M.O. Goerbig$^{2}$, and C. Morais Smith$^{1}$}
\affiliation{$^{1}$Institute for Theoretical Physics, Utrecht University, Leuvenlaan 4, 3584 CE Utrecht, The Netherlands}
\affiliation{$^{2}$Laboratoire de Physique des Solides, CNRS UMR 8502, Universit\'e Paris-Sud, 91405 Orsay, France}
\date{\today}

\begin{abstract}
Inspired by the recent creation of the honeycomb optical lattice and the realization of the Mott insulating state in a square lattice by shaking, we study here the shaken honeycomb optical lattice. For a periodic shaking of the lattice, a Floquet theory may be applied to derive a time-independent Hamiltonian. In this effective description, the hopping parameters are renormalized by a Bessel function, which depends on the shaking direction, amplitude and frequency. Consequently, the hopping parameters can vanish and even change sign, in an anisotropic manner, thus yielding different band structures. Here, we study the merging and the alignment of Dirac points and dimensional crossovers from the two dimensional system to one dimensional chains and zero dimensional dimers. We also consider next-nearest-neighbor hopping, which breaks the particle-hole symmetry and leads to a metallic phase when it becomes dominant over the nearest-neighbor hopping. Furthermore, we include weak repulsive on-site interactions and find the density profiles for different values of the hopping parameters and interactions, both in a homogeneous system and in the presence of a trapping potential. Our results may be experimentally observed by using momentum-resolved Raman spectroscopy.
\end{abstract}

\maketitle
\date{\today}

\section{Introduction}~

The study of Dirac points, i.e. the contact points between different energy bands with an approximate linear dispersion relation, has become a major issue since the experimental breakthrough in graphene-based electronics \cite{Novoselov04,Berger04}. Indeed, the low-energy electronic properties of graphene are governed by a pseudo-relativistic 2D Dirac equation for massless fermions situated at the $K$ and $K'$ corners of the Brillouin zone \cite{CastroNeto09}. The Dirac points are topologically protected and a gap is opened only when the inversion symmetry of the lattice or the time-reversal symmetry are broken.

The possibility to generate topological phase transitions in graphene-like systems has recently attracted a great deal of attention. Within a tight-binding description, an anisotropy in the nearest-neighbor hopping parameters makes the Dirac points move away from the high-symmetry $K$ and $K'$ points and, under appropriate conditions, merge at time-reversal invariant points in the first Brillouin zone \cite{Hasegawa06,Dietl,Montambaux09}. Most saliently, this merging of Dirac points is associated with a topological phase transition between a semimetallic phase and a gapped band-insulating phase. An experimental investigation of the merging transition in graphene turns out to be problematic, since in order to appropriately modify the hopping parameters, an unphysically large strain needs to be applied to the graphene sheet \cite{Strain0809}. 

An alternative system for the study of such topological transitions is that of ultracold atoms trapped in a honeycomb optical lattice. Since the seminal realization of the superfluid-Mott-insulator transition in the Bose-Hubbard model, ultracold atoms in optical lattices have become promising 
systems to emulate condensed-matter physics. Indeed, the lattice geometry, the dimensionality, the atomic species, as well as the interactions can be engineered with a high degree of precision. The more involved triangular and honeycomb geometries were recently realized experimentally and exotic correlated states of matter have been observed experimentally \cite{Honeycomb10} or predicted theoretically \cite{Sorella92,Meng10,Wang11,Bermudez10}.

The application of a time-periodic perturbation on the optical lattice introduces yet another parameter scale into the system. A periodic shaking of the optical lattice, up to the kHz frequency range, has been implemented by placing one of the mirrors used to create the optical lattice on a piezoelectric material, such that the mirror can be moved back and forth in the direction of the beam \cite{Lignier07,Zenesini09}. The Floquet formalism shows that the hopping energy of the atoms in the shaken lattice is renormalized by a Bessel function, as a function of the shaking frequency and amplitude, thus allowing both the magnitude and the sign of the hopping parameter to change. This rather counter-intuitive phenomenon, as compared to the standard tight-binding physics, has been experimentally observed in a one-dimensional cold-atomic system \cite{Zenesini09}.

In this paper, we consider ultracold fermions trapped in a \textit{shaken} honeycomb optical lattice. Within the Floquet formalism, we derive an effective Hamiltonian that generalizes that of a graphene-like material under strain. In particular, we find that the alignment and merging of Dirac points in momentum space are now accessible with ultracold fermions in the shaken optical lattice and the phase diagram consists of various phases of the corresponding solid-state system that are otherwise difficult to realize. Furthermore, by taking into account a Hubbard-like interaction for spinful fermions, we study the density profiles for the homogenous and the trapped gas within a Hartree-Fock theory.

The outline of this paper is the following: in Sec.\,\ref{sec01a} we introduce the time-dependent Hamiltonian and in Sec.\,\ref{sec01b} we derive the time-independent one, by applying the Floquet formalism. In Sec.\,\ref{sec02} we investigate the merging and alignment of Dirac points, when the optical lattice is shaken along specific directions. The description is extended to include interactions in Sec.\,\ref{sec03}, where we derive the dependence of the density on the chemical potential. Implications of our results for experiments are discussed in Sec.\,\ref{sec04}. Finally, our conclusions are presented in Sec.\,\ref{conc}.

\section{The shaken honeycomb lattice} \label{sec01}

In this section, we derive a time-independent effective description for ultracold atoms trapped in a periodically shaken honeycomb optical lattice by utilizing a Floquet theory. For simplicity, we focus on a system of single-component fermionic atoms and consider only single-particle terms in this section. The results from the Floquet theory are valid for fermionic atoms with internal degrees of freedom as well as for bosonic atoms. In particular, the hyperfine state of fermionic atoms, playing the role of an effective spin-$1/2$ degree of freedom for electrons, will be considered when interaction effects are taken into account in Sec. IV.

\subsection{Time-dependent Hamiltonian} \label{sec01a}
\begin{figure}[h]
\centering
\includegraphics[scale=0.3, angle=0]{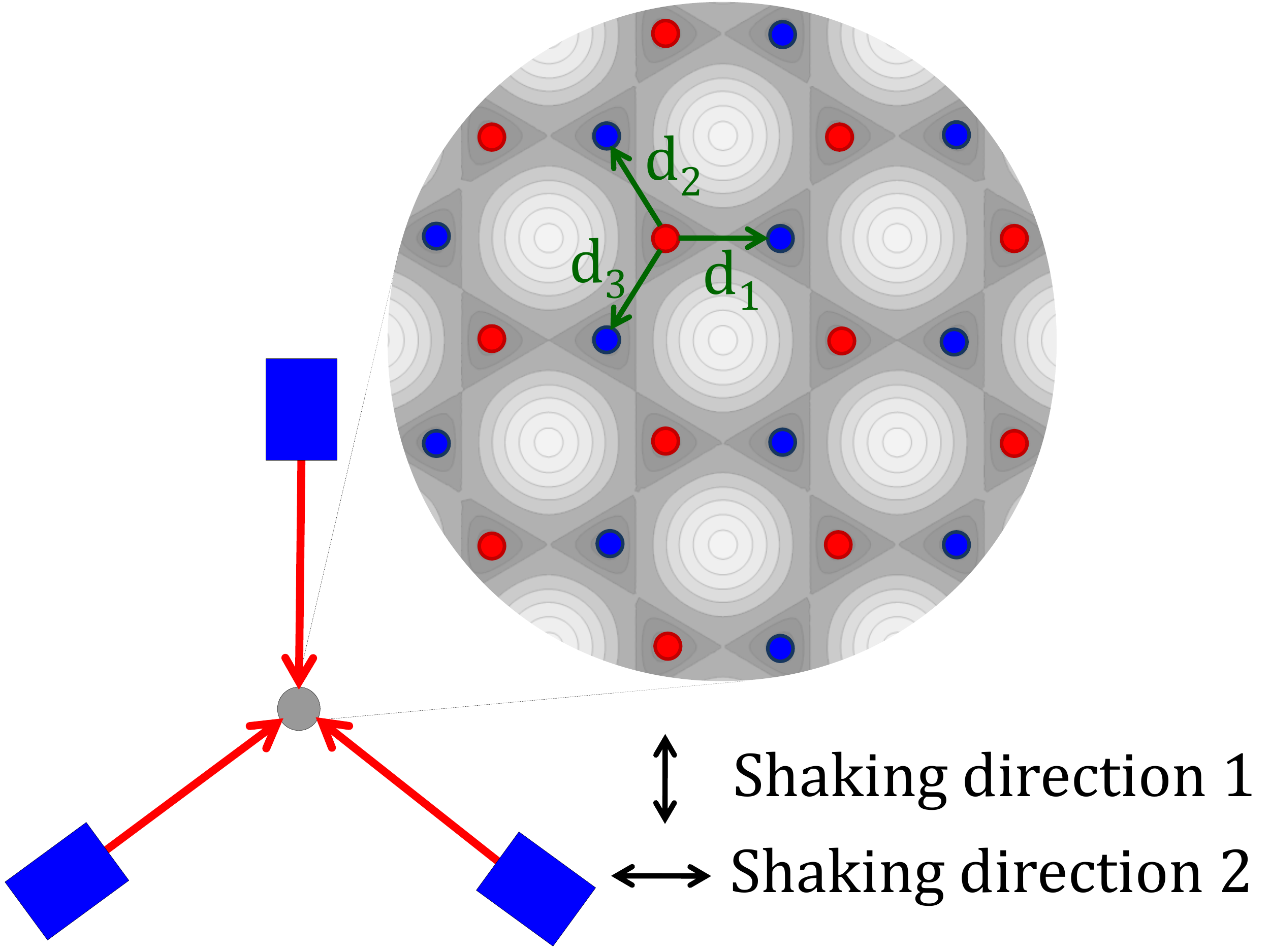}
\caption{(Color online) Laser configuration to create the honeycomb lattice, which consists of two triangular sublattices ($A$, red dots, and $B$, blue dots). The vectors $\textbf{d}_1$, $\textbf{d}_2$, and $\textbf{d}_3$ connect a site on the $A$ sublattice to its nearest neighbors on the $B$ sublattice.}
\label{fig:lattice}
\end{figure}

In the tight-binding limit, the system of ultracold fermionic atoms trapped in a 2D \textit{shaken} honeycomb optical lattice can be described by the Hamiltonian
\begin{equation}\label{basicH}
H(t) = H_0 + W(t) ,
\end{equation}
which consists of two distinct parts.
The static part
\begin{align}\label{eq:H0}
H_{0} = &- \gamma \sum_{j=1}^3 \sum_{\textbf{r} \in A} \left( a^\dagger_\textbf{r} b_{\textbf{r} + \textbf{d}_j} + b^\dagger_{\textbf{r} + \textbf{d}_j} a_\textbf{r} \right) \nonumber \\
&- \gamma' \sum^3_{i=1} \sum^3_{j=1,j\neq i} \bigg( \sum_{\textbf{r} \in A}  a^\dagger_\textbf{r} \, a_{\textbf{r} + \textbf{d}_i - \textbf{d}_j} + \sum_{\textbf{r} \in B} b^\dagger_\textbf{r} b_{\textbf{r} + \textbf{d}_i - \textbf{d}_j} \bigg) \nonumber \\
&- \mu \bigg( \sum_{\textbf{r} \in A} a^\dagger_\textbf{r} a_\textbf{r} + \sum_{\textbf{r} \in B} b^\dagger_\textbf{r} b_\textbf{r} \bigg)
\end{align}
is simply the tight-binding Hamiltonian in the honeycomb lattice, 
where
$a^\dagger_\textbf{r}$ ($b^\dagger_\textbf{r}$) and $a_\textbf{r}$ ($b_\textbf{r}$) are, respectively, fermionic creation and annihilation operators on the lattice site $\textbf{r}$ in the $A$ ($B$) sublattice. The three vectors
\begin{equation}
\textbf{d}_1=d \hat{e}_x, \,\textbf{d}_2=\frac{d}{2}(-\hat{e}_x+\sqrt{3}\hat{e}_y), \,\textbf{d}_3=\frac{d}{2}(-\hat{e}_x-\sqrt{3}\hat{e}_y),
\label{eq:nn-vec}
\end{equation}
connect an $A$-lattice site with its three nearest-neighbor (nn) $B$-lattice sites and are given in terms of the distance $d=8\pi/3\sqrt{3}k$ between nn sites, where $k$ is the laser wave number (see Fig.\,\ref{fig:lattice}). Here, $\gamma,\gamma'>0$ characterize the energy gained in hopping to the nn and next-nearest-neighbor (nnn) sites, respectively, and $\mu$ is the on-site energy. We remark that the nnn hopping is taken into account because the nn hopping may be rendered vanishingly small in the effective time-independent description. In this regime, the nnn may become the dominant kinetic term. In a square lattice, where the potential is separable in independent $\hat{e}_x$ and $\hat{e}_y$ components, the nnn hopping is identically zero \cite{Liberto11}. However, the nnn hopping can be nonzero in the honeycomb lattice, since its potential is not separable in $\hat{e}_x$ and $\hat{e}_y$ components. Nevertheless, it may be expressed as the sum of two triangular lattices.

The time-dependent part of the Hamiltonian (\ref{basicH}),
\begin{equation}
W(t) = m \Omega^2 \cos(\Omega t) \left( \sum_{\textbf{r} \in A} \textbf{r}\cdot\boldsymbol{\rho} \, a^\dagger_\textbf{r} a_\textbf{r} + \sum_{\textbf{r} \in B} \textbf{r}\cdot\boldsymbol{\rho} \, b^\dagger_\textbf{r} b_\textbf{r} \right),
\end{equation}
describes the harmonic shaking of the lattice in the direction $\boldsymbol{\rho}$ with a driving frequency $\Omega$ in the co-moving frame of Ref. \cite{Gluck02}. As a consequence of the transformation to the co-moving frame, $W(t)$ describes atoms of mass $m$ experiencing a position-dependent sinusoidal force.

\subsection{Effective Hamiltonian} \label{sec01b}
The unavoidable complexity that arises when dealing with a quantum many-body system out of equilibrium has recently motivated the development of new theoretical tools, for example time-dependent density matrix renormalization group \cite{Rengroup0405}, time-dependent dynamical mean field theory \cite{Dynmeanfield0609}, and exact diagonalization \cite{Rigol09}.

However, for a periodically driven quantum system, the Floquet theory offers a simplified description of the system, in the form of a time-independent effective Hamiltonian, if the period $T=2\pi/\Omega$ is the shortest time scale in the problem \cite{Grifoni98}. In this limit, the atoms cannot follow the shaking motion adiabatically and remain thus at their average lattice position, albeit with renormalized hopping parameters. The system is thus considered to be in a stationary state and the knowledge of equilibrium physics can be employed.

Let us consider the Floquet Hamiltonian defined by $H_F = H(t) - i \hbar \partial_t$, where $H(t+T)=H(t)$ is periodic in time \cite{Grifoni98}. The eigenvalue equation is then given by
\begin{equation}
H_F |\phi (q,t) \rangle = \epsilon_{\phi} | \, \phi (q,t) \rangle,
\label{eq:floquet-schrodinger}
\end{equation}
where $\epsilon_{\phi}$ is the quasienergy defined uniquely up to a multiple of $\hbar \Omega$. Any solution $| \, \phi (q,t) \rangle$ is part of a set of solutions $ \exp(i n \Omega t) | \, \phi (q,t) \rangle$ with integer $n$, which all correspond to the same physical solution. Hence, the spectrum of the Floquet Hamiltonian possesses a Brillouin-zone-like structure \cite{Grifoni98}. The interest therefore lies with the states in the first Brillouin zone, i.e. states with quasi-energies $ - \hbar \Omega /2 < \epsilon_\phi \leq \hbar \Omega /2 $.

The space in which the states $| \, \phi (q,t) \rangle$ are defined is the composite of the Hilbert space spanned by square integrable functions on configuration space, $|\alpha (q) \rangle$,  and the space of $T$-periodic functions. The state $| \, \phi (q,t) \rangle$ may be written down in an orthonormal basis in the composite space according to
\begin{equation}
| \phi (q,t) \rangle = \sum_{n=0}^\infty \sum_{\alpha} c_{n,\alpha} \exp[-i \hat{F}(t)+ i n \Omega t ] |\alpha (q) \rangle,
\end{equation}
where $c_{n,\alpha}$ are coefficients to normalize $| \, \phi (q,t) \rangle$ and the operator $\hat{F}(t)$ can be any $T$-periodic Hermitian operator. Therefore, we can conveniently choose $\hat{F}(t)$ to be $\hat{F}(t) = \hbar^{-1} \int_0^t dt' W(t')$, such that $ H(t) - \hbar \partial_t \hat{F}(t) = H_0$.
\begin{widetext}
If the condition
\begin{equation}
\langle \alpha'(q) | \langle \exp[i \hat{F}(t) ] \, \exp[ i (n-n') \Omega t] \, H_0 \exp[-i \hat{F}(t) ] \rangle_T | \, \alpha (q) \rangle \ll \hbar \Omega
\label{eq:condition}
\end{equation}
is satisfied for any two states $| \, \alpha (q) \rangle$ and $| \, \alpha' (q) \rangle$, then the eigenvalues $\epsilon_\phi$ are approximately
\begin{equation}
\epsilon_\phi = \langle \phi (q,t) | \langle H_F\rangle_T | \phi (q,t) \rangle \approx \sum_{\alpha, \alpha'} c_{0,\alpha'} c_{0,\alpha}  \langle \alpha'(q) | \, \langle \exp[i \hat{F}(t) ] \, H_0  \, \exp[-i \hat{F}(t) ] \rangle_T  \,| \,\alpha (q) \rangle .
\label{eq:floquet-approximation}
\end{equation}
\end{widetext}
Here, $\langle \mathcal{O}(t) \rangle_T=T^{-1}\int_0^{T}dt\, \mathcal{O}(t)$ denotes the time average of the operator $\mathcal{O}(t)$ over the period $T$. The condition ($\ref{eq:condition}$) will hold for $n \neq n'$, if $H_0$ is nearly constant during the period $T$, which is small if $\Omega$ is large. In this case, states with different $n$ do not mix. If $\Omega$ is large enough, such that the condition ($\ref{eq:condition}$) also holds for $n=n'$, then the energy spectrum will split up into energy bands labelled by an index $n$, where the details within the energy band are determined by $H_0$. Because the states with a different index $n$ are separated by an energy which is a multiple of $\hbar \Omega$ and because the spectrum possesses a Brillouin-zone-like structure, only the terms with $n=0$ need to be taken into account. The effective Hamiltonian $H_\textrm{eff}$, which gives rise to the same spectrum as the Floquet Hamiltonian, is then defined by \cite{Hemmerich10}
\begin{align}
H_\textrm{eff} &= \bigg< \exp[i \hat{F}(t) ] \, H_{0} \, \exp[-i \hat{F}(t) ] \bigg>_T \nonumber \\
&= \bigg< \sum^\infty_{n=0} \frac{i^n}{n !} [\hat{F}(t), H_{0}]_n \bigg>_T.
\label{eq:effective-ham}
\end{align}
Here, $[\hat{F},\hat{G}]_n$ denotes the multiple commutator, which is defined by $[\hat{F},\hat{G}]_{n+1} = [\hat{F},[\hat{F},\hat{G}]_n]$ and $[\hat{F},\hat{G}]_0 = \hat{G}$.

Effective Hamiltonians corresponding to Eq.\,(\ref{eq:effective-ham}) have been derived for linear shaking of a one-dimensional lattice \cite{Eckardt05}  and for elliptical shaking of a triangular lattice \cite{Eckardt10}. For the shaken honeycomb lattice studied here, the condition ($\ref{eq:condition}$) is satisfied if $\gamma \ll \hbar \Omega$, and  the effective Hamiltonian becomes 
\begin{align}
H_{\textrm{eff}} = &- \sum_{j=1}^3 \sum_{\textbf{r} \in A} \gamma_j \left( a^\dagger_\textbf{r} b_{\textbf{r} + \textbf{d}_j} + b^\dagger_{\textbf{r} + \textbf{d}_j} a_\textbf{r} \right) \nonumber \\
&- \sum^3_{i=1} \sum^3_{j=1,j\neq i} \gamma'_{i,j} \bigg( \sum_{\textbf{r} \in A} a^\dagger_\textbf{r} \, a_{\textbf{r} + \textbf{d}_i - \textbf{d}_j}
+ \sum_{\textbf{r} \in B} b^\dagger_\textbf{r} b_{\textbf{r} + \textbf{d}_i - \textbf{d}_j} \bigg) \nonumber \\
&- \mu \bigg( \sum_{\textbf{r} \in A} a^\dagger_\textbf{r} a_\textbf{r} + \sum_{\textbf{r} \in B} b^\dagger_\textbf{r} b_\textbf{r} \bigg),
\label{Hamiltonian-eff}
\end{align}
where the renormalized nn hopping parameters $\gamma_j$ are given by
\begin{equation}
\gamma_j = \gamma J_0 \left( \bigg| \textbf{d}_j \cdot \boldsymbol{\rho} \frac{m \Omega}{\hbar} \bigg| \right) ,
\label{eq:renormalized-gamma}
\end{equation}
and the renormalized nnn hopping parameters are given by
\begin{equation}
\gamma'_{i,j} = \gamma' J_0 \left( \bigg| (\textbf{d}_i - \textbf{d}_j) \cdot \boldsymbol{\rho} \frac{m \Omega}{\hbar} \bigg| \right)
\label{eq:renormalized-gammaprime}
\end{equation}
(see Appendix \ref{app1} for detailed calculations). In these expressions, $J_0(x)$ denotes the zeroth order Bessel function of the first kind, which shows a damped oscillation around zero.

In terms of the renormalized nn and nnn hopping parameters, the diaginalization of the effective Hamiltonian (\ref{Hamiltonian-eff}) 
yields the dispersion relation
\begin{equation}
\label{DispersionRel} 
\epsilon_{\lambda}(\bq)=h(\bq)+\lambda|f(\bq)|,
\end{equation}
where $\lambda=\pm$ is the band index, and we have defined the functions
\begin{equation}\label{fq}
f(\textbf{q}) = \sum_j \gamma_j \exp(-i \textbf{q} \cdot \textbf{d}_j)
\end{equation}
and 
\begin{equation}
h(\bq)=2\sum_{i<j}\gamma_{i,j}^{\prime}\cos\left[\bq\cdot(\textbf{d}_i-\textbf{d}_j)\right].
\end{equation}

\section{Merging and alignment of Dirac points} \label{sec02}

In this section, the honeycomb lattice with anisotropic hopping is studied. In the first two subsections, only nn hopping is considered
for illustration reasons. Indeed, this allows for a simple understanding of the main consequences of shaking on
Dirac-point motion and dimensional
crossover. In Subsec. \ref{sec02c}, we discuss how the picture evolves when nnn hopping is included, and the sign of the hopping
parameters is investigated in Subsec. \ref{sec02d}.
Since the system with two nn equal hopping parameters and a single independent one captures the essential features of the systems with three independent nn hopping parameters, we will focus on this system. The numbering of the $\gamma_j$s is chosen such that $|\gamma_2| = |\gamma_3| = \gamma_{2,3}$, which can be achieved by shaking in a direction parallel or perpendicular to $\textbf{d}_1$. 

Although the atoms in the optical lattice are charge-neutral objects, we shall adopt the language from condensed-matter physics and call a zero-gap phase with a pair of Dirac cones and a vanishing density of states at the band-contact points a \textit{semimetal}, whereas a gapped phase is called \textit{band insulator}. Furthermore, nnn hopping induces a \textit{metallic} phase for small values of $\gamma_j$ because of an overlap between the two bands that yields a non-vanishing density of states at the energy level of the band-contact points.

\subsection{Merging of Dirac points} \label{sec02a}
If the latice is shaken in the direction perpendicular to $\textbf{d}_1$ (direction 1 in Fig.\,\ref{fig:lattice}), $\gamma_1$ remains equal to $\gamma$, whereas $\gamma_2$ and $\gamma_3$ are renormalised to a smaller value. An increase in the shaking amplitude results in a decrease in $\gamma_{2,3} = \gamma_2 = \gamma_3$, which is depicted by the arrow $M$ in Fig.\,\ref{fig:phase-diagram-abs}(a). When the hopping parameters change according to this arrow $M$, the energy spectrum evolves from Fig.\,\ref{fig:merging-energy-dispersion}(a) to Fig.\,\ref{fig:merging-energy-dispersion}(b). The Dirac points, originally situated at the corners $K$ and $K'$ of the first Brillouin zone, start to move in the $q_y$-direction along the vertical edges of the latter. This motion is depicted by the arrows in Fig.\,\ref{fig:phase-diagram-abs}(b). Even if the two Dirac points are no longer located at the high-symmetry points $K$ and $K'$, they remain related by time-reversal symmetry, such that their Berry phases $\pi$ and $-\pi$ are opposite. This non-zero Berry phase topologically protects each of the Dirac points and thus the semimetallic phase remains robust until $\gamma_{2,3}=\gamma_1/2$, where the two points merge at a time-reversal invariant momentum, i.e. half of a reciprocal lattice vector \cite{Montambaux09}. In the present example, this point is situated at the center of the vertical edges of the first Brillouin zone, and the band dispersion becomes parabolic in the $y$-direction while remaining linear in the $x$-direction [see Fig.\,\ref{fig:merging-energy-dispersion}(b)]. The merged Dirac points are no longer topologically protected due to the annihilation of the opposite Berry phases. Consequently, a further increase of the shaking amplitude, which results in a further decrease of $\gamma_{2,3}$, leads to the opening of a gap between the two bands. Thus, the system undergoes a topological phase transition from a semimetal to a band insulator. This merging transition was also studied in a static setup in Ref. \cite{Lee09}, where the hopping amplitudes $\gamma$ were proposed to be modified by a change in the intensity of one of the lasers used to create the optical lattice. In contrast to this static setup, shaking the honeycomb lattice allows one to completely annihilate some of the nn hopping parameters and to even change their sign. This sign change occurs at the zeros of the Bessel function [see Eq.\,(\ref{eq:renormalized-gamma})]. For an example system of $^{40}$K atoms in a lattice created by lasers with a wavelength of 830 nm, which is shaken in the direction perpendicular to $\textbf{d}_1$, the situation $\gamma_{2,3}=0$ is encountered for
{\begin{equation}
\rho = 180 \textrm{nm}; \; \Omega/2\pi = 6 \textrm{kHz},
\label{eq:dim-cross-over-0d}
\end{equation}
which corresponds to the first zero of the Bessel function. At this particular point, and if $\gamma'=0$ in addition, the system consists of a set of effectively decoupled horizontal bonds along which the atoms are solely allowed to hop. This yields two flat bands at $\pm \gamma_1$ [see Fig.\,\ref{fig:merging-energy-dispersion}(c)] that may be viewed as the extreme limit of the band-insulating phase. Alternatively, one may view this situation upon decreasing the value of $\gamma_{2,3}$ as a dimensional crossover from a 2D band insulator to a zero-dimensional (0D) system. A small non-zero value of $\gamma_{2,3}$ simply provides a weak dispersion of these decoupled bands (not shown).

\begin{figure}[h]
\begin{minipage}{4cm}
	\centering
		\includegraphics[scale=0.16, angle=0]{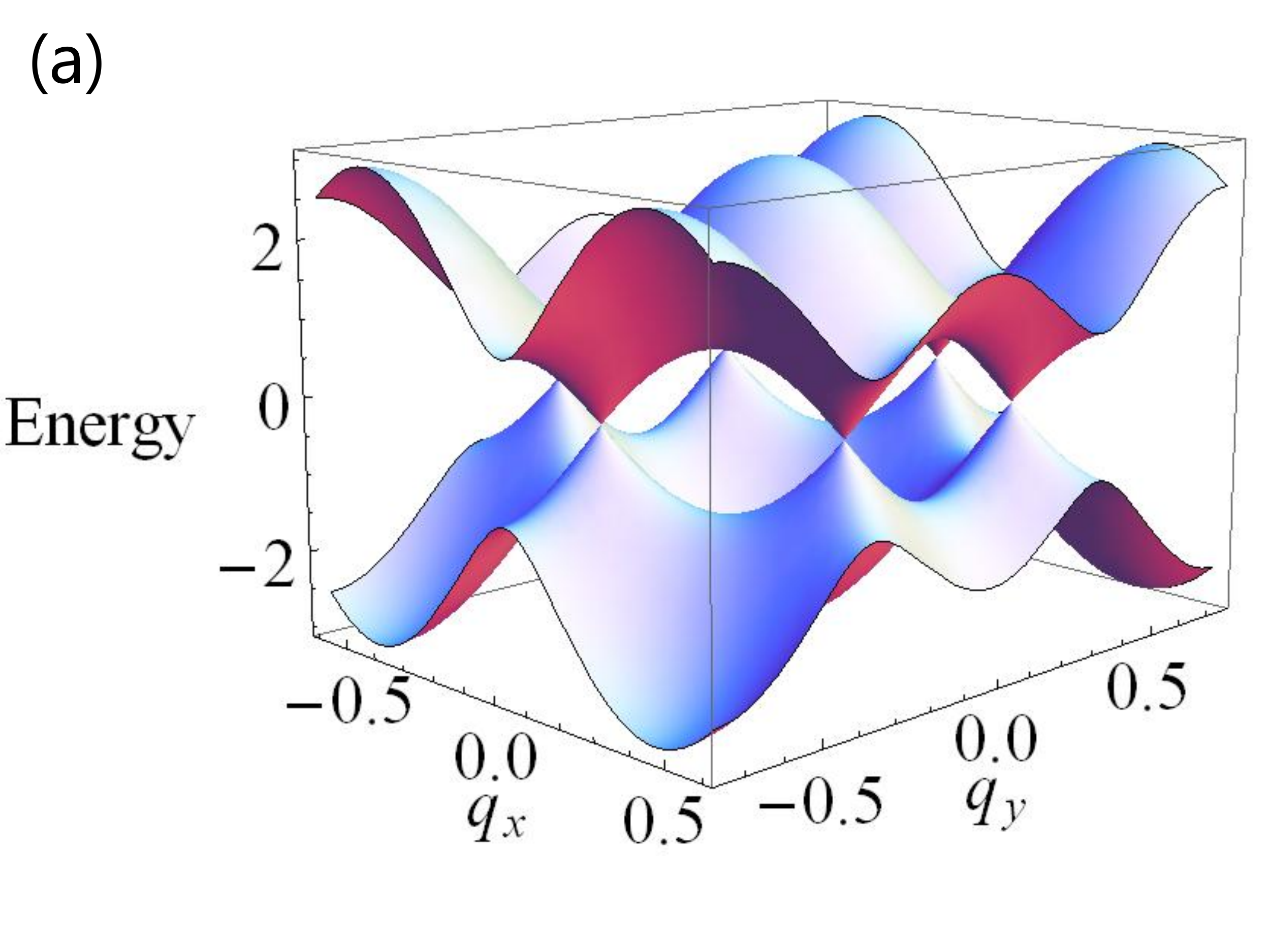}
\end{minipage}
\begin{minipage}{4cm}
	\centering
		\includegraphics[scale=0.16, angle=0]{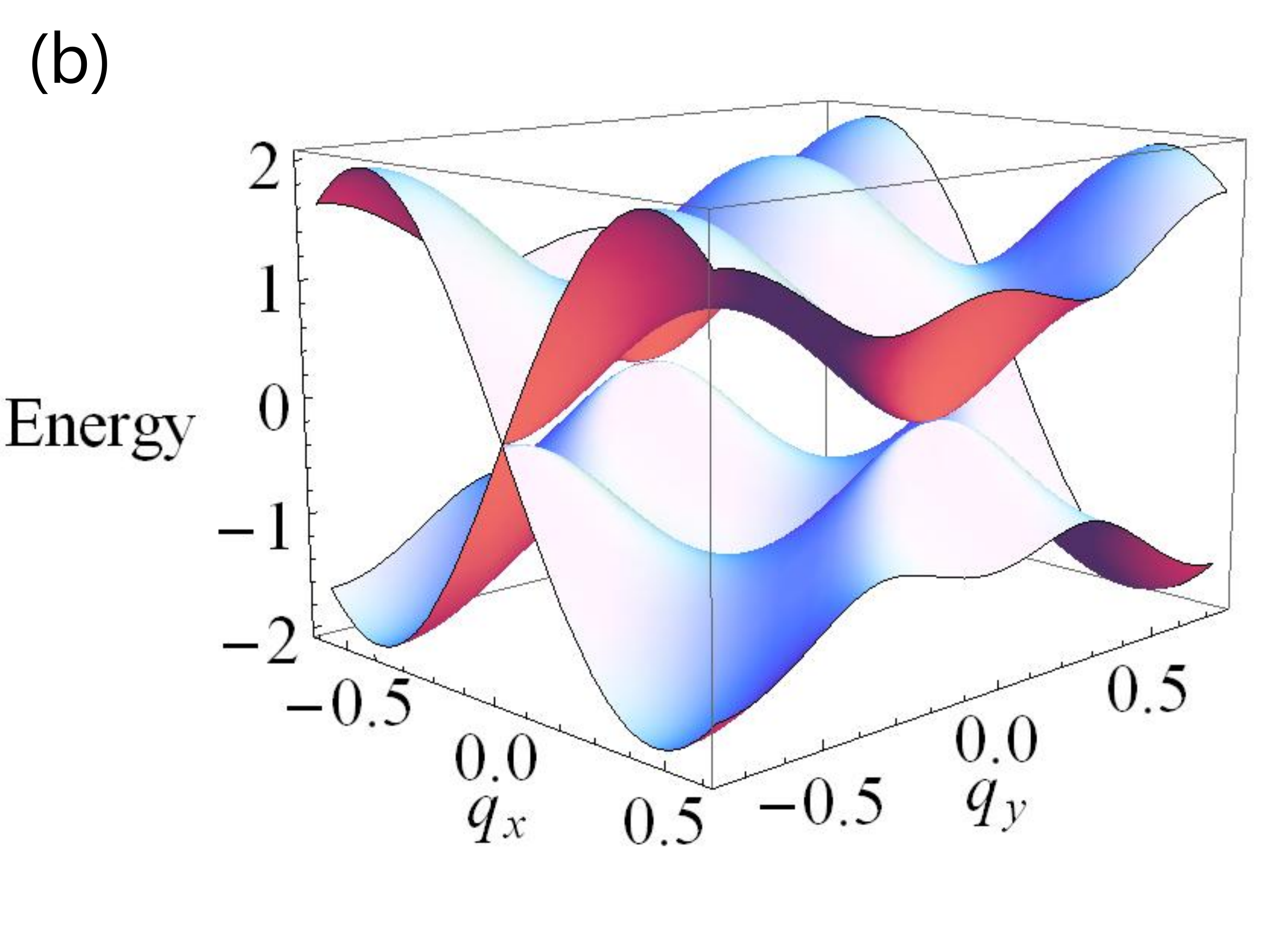}
\end{minipage}

\begin{minipage}{4cm}
	\centering
		\includegraphics[scale=0.16, angle=0]{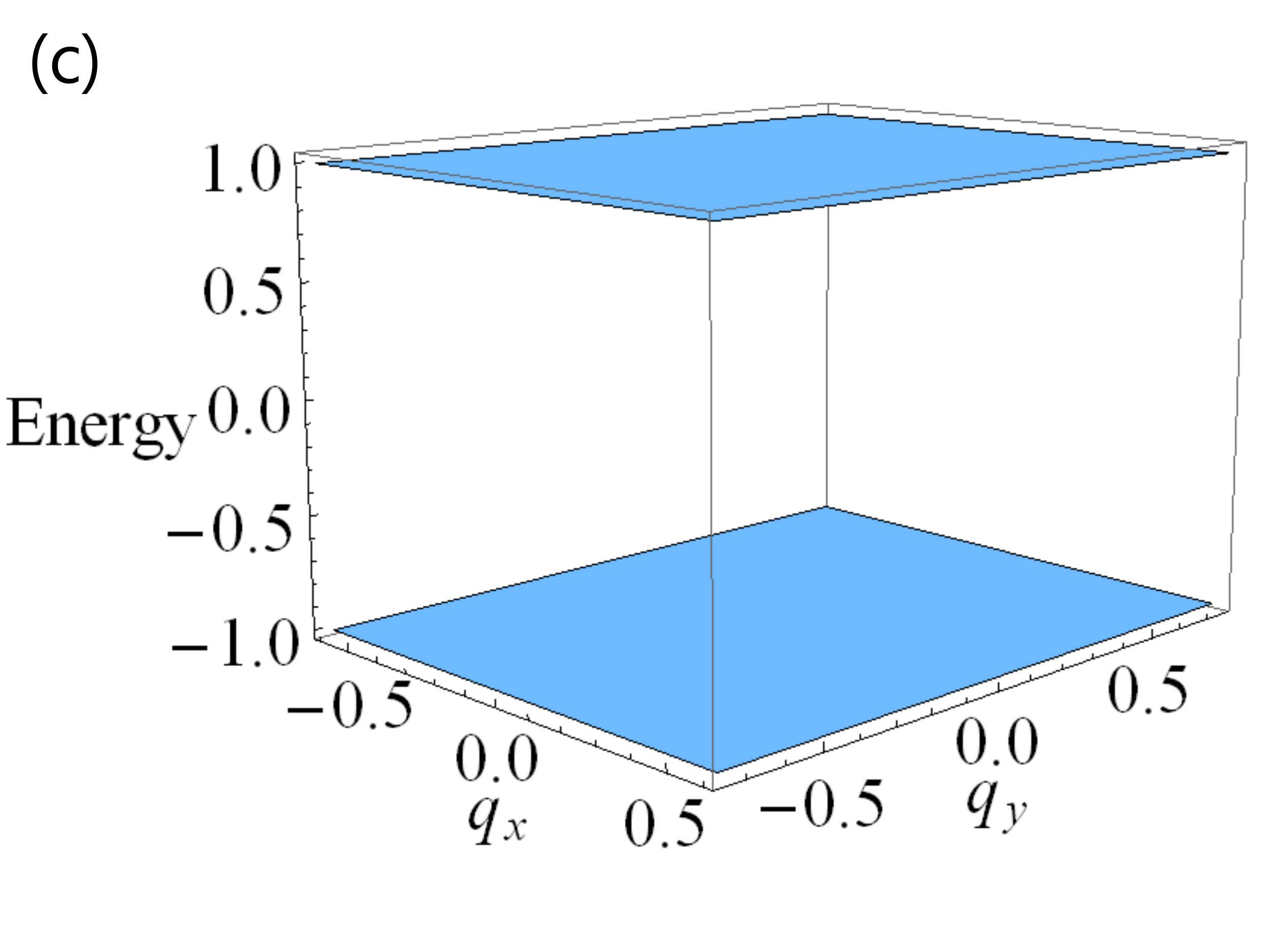}
\end{minipage}
\begin{minipage}{4cm}
	\centering
		\includegraphics[scale=0.16, angle=0]{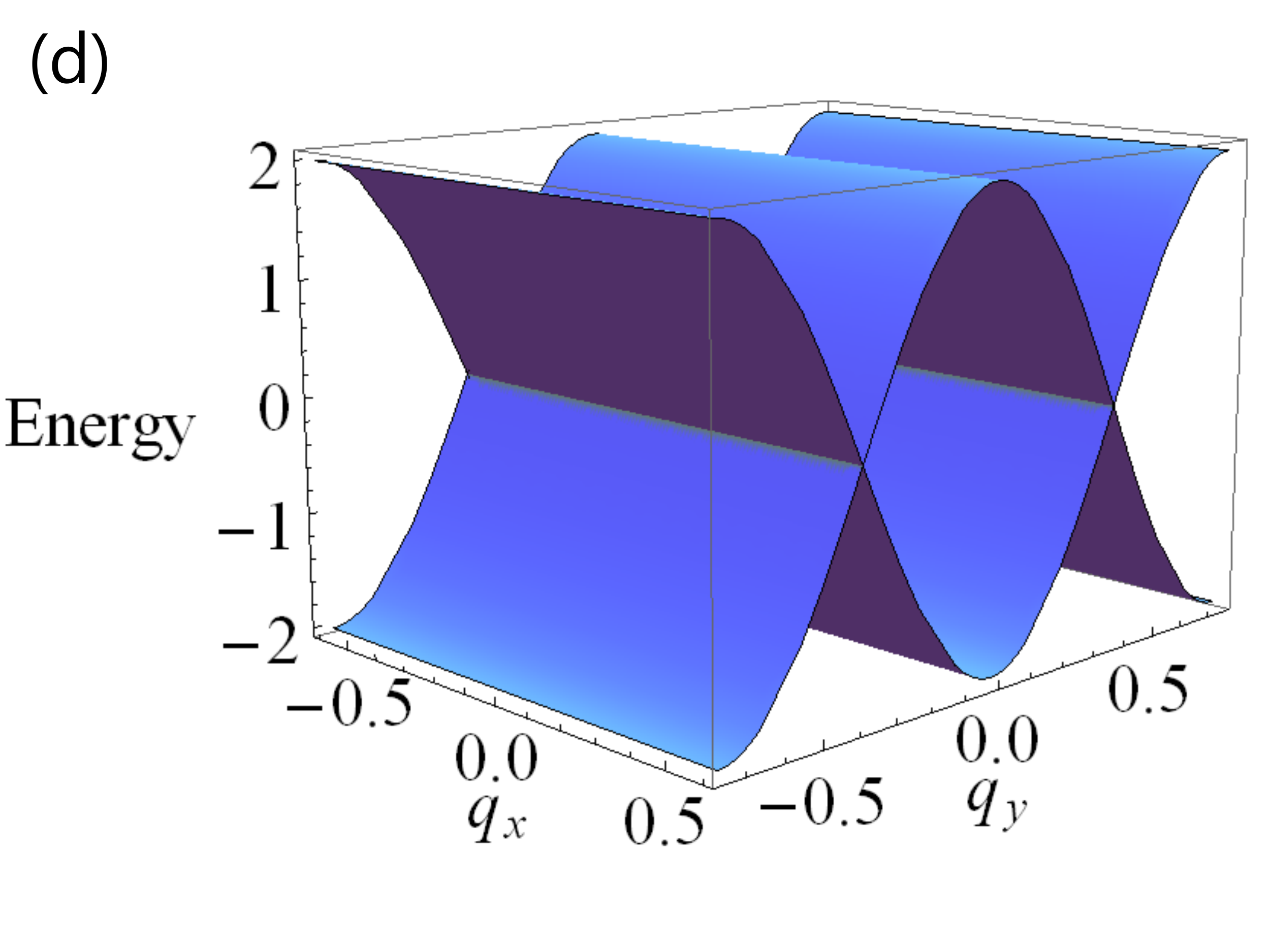}
\end{minipage}

\caption{(Color online) Energy dispersion for the shaken honeycomb optical lattice, with $k=1$, $\gamma =1$, and $\gamma'=0$. The labels $q_x$ and $q_y$ represent the $x$ and $y$ components of the momentum, respectively. The $x$ and $y$ axes have been chosen such that the nn vectors are given by Eq.\,(\ref{eq:nn-vec}). (a) The isotropic case, where $\gamma_1=\gamma_{2,3}$. (b) The merged Dirac points, where $\gamma_{2,3} = \gamma_1/2$. (c) The zero dimensional case, where $\gamma_{2,3}=0$. (d) The aligned Dirac points, where $\gamma_1=0$.}
\label{fig:merging-energy-dispersion}
\end{figure}

\begin{figure}[h]
	\centering
		\includegraphics[scale=0.3, angle=0]{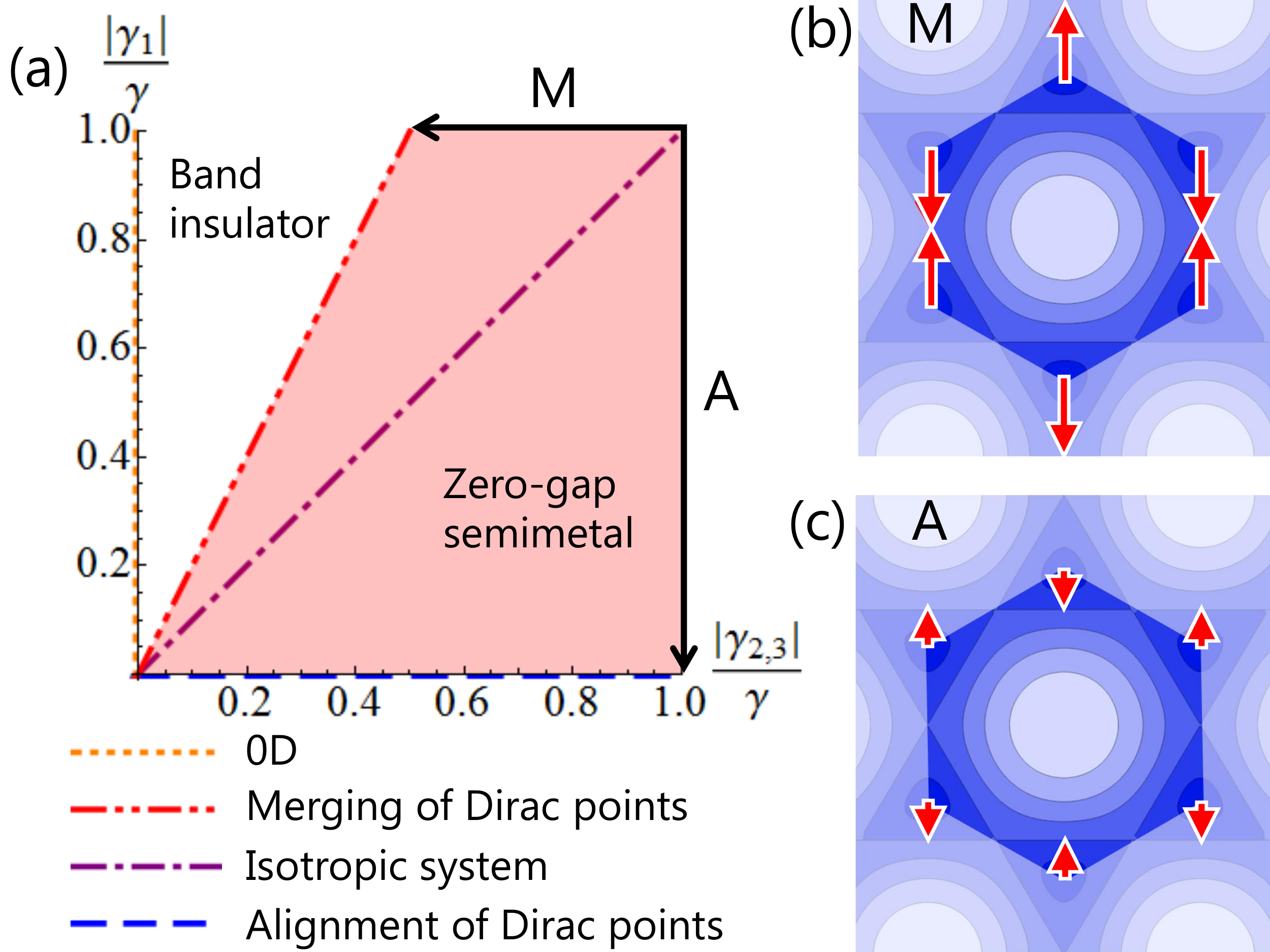}
	\caption{(Color online) (a) Phase diagram, showing the phase transition between the zero-gap semimetallic phase and the insulating phase, which happens at $\gamma_1 = 2 \gamma_{2,3}$. Here, we have chosen $\gamma'=0$. (b) Dirac-point motion in the first Brillouin zone for a shaking direction perpendicular to $\textbf{d}_1$ [direction $M$ in the phase diagram (a)]. (c) Dirac-point motion in the first Brillouin zone for a shaking direction parallel to $\textbf{d}_1$ [direction $A$ in the phase diagram (a)]. The contour plots depict the dispersion of the isotropic system with an arbitrary color scale. The area with higher contrast is the first Brillouin zone.}
	\label{fig:phase-diagram-abs}
\end{figure}

\subsection{Alignment of Dirac points} \label{sec02b}
Another dimensional crossover, from 2D to 1D, may be obtained if the lattice is shaken in the direction parallel to one of the nn vectors (direction 2 in Fig.\,\ref{fig:lattice}). Here, we choose ${\bf d}_1$ to maintain the symmetry $\gamma_{2,3}=\gamma_2=\gamma_3$. In this case, both $\gamma_1$ and $\gamma_{2,3}$ are renormalized by Bessel functions, albeit with different arguments. Since all hopping parameters are renormalized, the trajectory of the system in the phase space upon increasing the shaking amplitude is not a straight line, as was the case for shaking perpendicular to a hopping direction, and has a new feature: the alignment of Dirac points, which occurs for $\gamma_1 = 0$. The first zero of $\gamma_1$ is found at
\begin{equation}
\rho = 92 \textrm{nm}; \; \Omega/2\pi = 6 \textrm{kHz},
\label{eq:dim-cross-over-1d}
\end{equation}
for the same system of $^{40}$K atoms mentioned above. Here, for illustrative purposes, a simplified trajectory of the system is depicted in Fig.\,\ref{fig:merging-energy-dispersion} by arrow $A$, which corresponds to the motion of the Dirac points in reciprocal space as shown in Fig.\,\ref{fig:phase-diagram-abs}(c). As $\gamma_1$ approaches zero, the Dirac points align in lines parallel to the $x$-axis at $q_y=\pm \pi/\sqrt{3}d$ and the energy barriers between the aligning points are lowered. Consequently, when $\gamma_1 = 0$, the energy spectrum contains lines where the two energy band meet and the dispersion is linear, as is shown in Fig.\,\ref{fig:merging-energy-dispersion}(d). The dispersion relation (\ref{DispersionRel}) then reads simply 
\begin{equation}
\epsilon_{\lambda}(\bq)=2\lambda \gamma_{2,3}\left|\cos\left(\frac{\sqrt{3}}{2}q_yd\right)\right|,
\end{equation}
and one clearly sees the 1D character. Indeed, there is no dispersion in the $q_x$-direction, as is also evident from 
Fig.\,\ref{fig:merging-energy-dispersion}(d), and the system may be viewed as completely decoupled 1D chains in which the zig-zag 
arrangement is of no importance. In this particular limit, the sites A and B are therefore no longer inequivalent such that the
unit cell is effectively divided by two, and the size of the first Brillouin zone is consequently doubled. The aligned Dirac points
may thus, alternatively, be viewed as due to an artificial folding of the second (outer) half of the first Brillouin zone into its 
inner half. However, this aspect is very particular in that the Brillouin zone immediately retrieves its original size when
$\gamma_1$ is small, but non-zero, or if nnn hoppings are taken into account. In both cases, one needs to distinguish the two
different sublattices and one obtains a dispersion in the $q_x$-direction.

 The actual behavior of the system for an increasing shaking amplitude is discussed in Sec.\,\ref{sec04}. This behavior is more complicated because all three nn hopping parameters are renormalized, which, beyond the alignment, leads to the merging of Dirac points and the opening of a gap also in the case of shaking parallel to one of the nn vectors. In the absence of nnn hopping, the 0D limit can be reached in addition.

\subsection{Next-nearest-neighbor hopping} \label{sec02c}

The major consequence of nnn hopping is to break particle-hole symmetry, as may be seen from Eq. (\ref{DispersionRel}), where non-zero
values of $\gamma_{i,j}^{\prime}$ yield $\epsilon_{\lambda}(\bq)\neq -\epsilon_{-\lambda}(\bq)$. Its relevance depends sensitively on the
shaking direction, because of the different renormalization of the nn hopping parameters. The band structure with nnn hopping included
is depicted in Fig.\,\ref{fig:nnn-energy-dispersion} for different shaking directions.

\subsubsection{Shaking in the direction perpendicular to $\textbf{d}_1$}

In the case of a shaking perpendicular to $\textbf{d}_1$, only $\gamma_{2,3}$ are decreased, whereas $\gamma_1=\gamma$ remains 
the leading energy scale in the band structure \footnote{We concentrate on $\textbf{d}_1$ as a reference direction, but it 
may naturally be replaced by any other direction $\textbf{d}_j$, in which case $\gamma_j=\gamma$ remains constant.}.
The band structure for the unshaken lattice is depicted in Fig.\,\ref{fig:nnn-energy-dispersion}(a) for 
$\gamma'/\gamma=0.1$, and one notices that the main features of the band structure, namely the Dirac points, are unaltered 
with respect to the case $\gamma'=0$ in Fig.\,\ref{fig:merging-energy-dispersion}(a), apart from the flattening of the upper band
as compared to the lower one. When approaching the merging transition $\gamma_{2,3}=\gamma_1/2$, the value of which is determined
by the zeros of $f(\bq)$ in Eq. (\ref{fq}) and that therefore does not depend on the nnn hopping parameters, the band width 
remains dominated by the largest hopping parameter $\gamma_1$, such that the band structure [Fig.\,\ref{fig:nnn-energy-dispersion}(b)]
at the transition is essentially the same as in Fig.\,\ref{fig:merging-energy-dispersion}(b) for $\gamma'=0$. In the 0D limit,
with $\gamma_{2,3} = 0$ the originally flat bands [Fig.\,\ref{fig:merging-energy-dispersion}(c)] acquire the weak dispersion 
of a triangular lattice as a 
consequence of the non-zero nnn hopping parameters. However, as expected from the above arguments, the dispersion is on the 
order of $\gamma'$ and thus small as compared to the energy separation $\sim 2\gamma_1=2\gamma$ between the two bands.

\subsubsection{Shaking in the direction parallel to $\textbf{d}_1$}

In contrast to a shaking direction perpendicular to $\textbf{d}_1$, nnn hopping has more drastic consequences if the lattice is
shaken in the direction parallel to $\textbf{d}_1$. In this case, all nn hopping parameters are decreased, and the relative 
importance of nnn hopping is enhanced. Notice further that the nnn lattice vectors $\pm(\textbf{d}_2-\textbf{d}_3)$ are now
perpendicular to the shaking direction such that $\gamma_{2,3}^{\prime}=\gamma'=0.1\gamma$ remains unrenormalized.
Also in this case, the system is approaching the 1D limit, with $\gamma_{1}=0$
[see Fig.\,\ref{fig:nnn-energy-dispersion}(d)]. However, in contrast to Fig.\,\ref{fig:merging-energy-dispersion}(d), 
the chains remain coupled by nnn hopping that yields a dispersion in the $q_x$-direction. Furthermore, as mentioned above,
the A and B sites are now not equivalent from a crystallographic point of view, such that the outer parts of the first 
Brillouin zone cannot be folded back into the inner one, as may be seen from Fig.\,\ref{fig:nnn-energy-dispersion}(d).

Finally, for particular values of the shaking amplitude in the direction parallel to $\textbf{d}_1$, the nn 
hopping parameters can be decreased in such a manner as to render $\gamma_{2,3}$ more relevant. In this case, the
two bands can overlap in energy, as depicted in Figs.\,\ref{fig:nnn-energy-dispersion}(e) and \,\ref{fig:nnn-energy-dispersion}(f) for
$\boldsymbol{\rho} = 5.2 (\hbar / m \Omega d) \hat{e}_x$ 
(in which case $\gamma_1 \approx \gamma_{2,3}$) and $\boldsymbol{\rho} = 4.8 (\hbar / m \Omega d) \hat{e}_x$
(with $\gamma_{2,3} \approx 0$),
respectively. In the latter example there are no band contact points, in spite of the overlap between the two bands, and
the system would be in an insulating phase if nnn hopping terms were not taken into account. This overlap in energy between the 
two bands yields a non-zero density of states at any energy, such that the semi-metallic (or insulating) phase vanishes and
yields, at half-filling, a metallic phase with particle and anti-particle pockets in the first Brillouin zone.}

\begin{figure}[h]
\begin{minipage}{4cm}
	\centering
		\includegraphics[scale=0.16, angle=0]{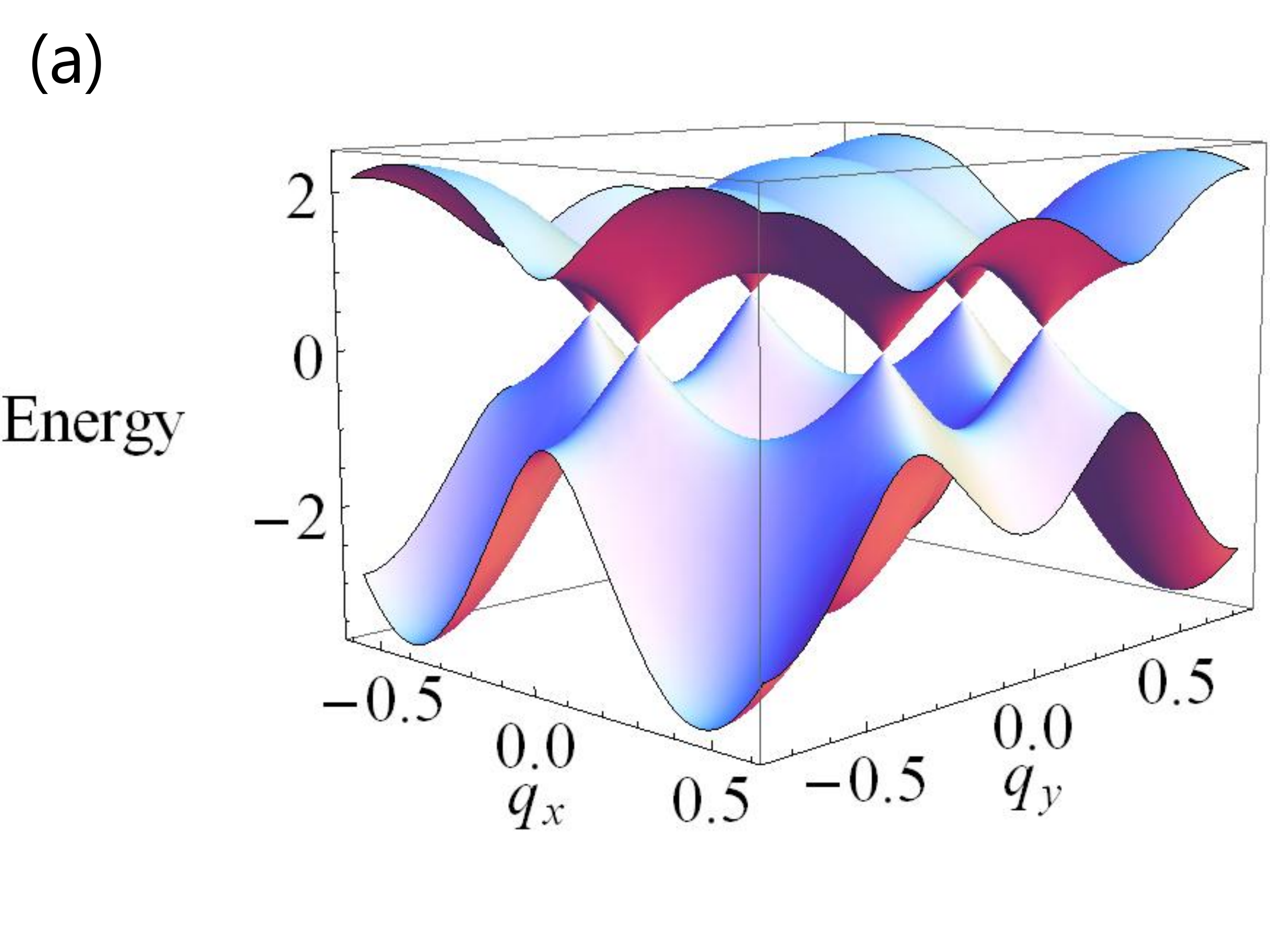}
\end{minipage}
\begin{minipage}{4cm}
	\centering
		\includegraphics[scale=0.16, angle=0]{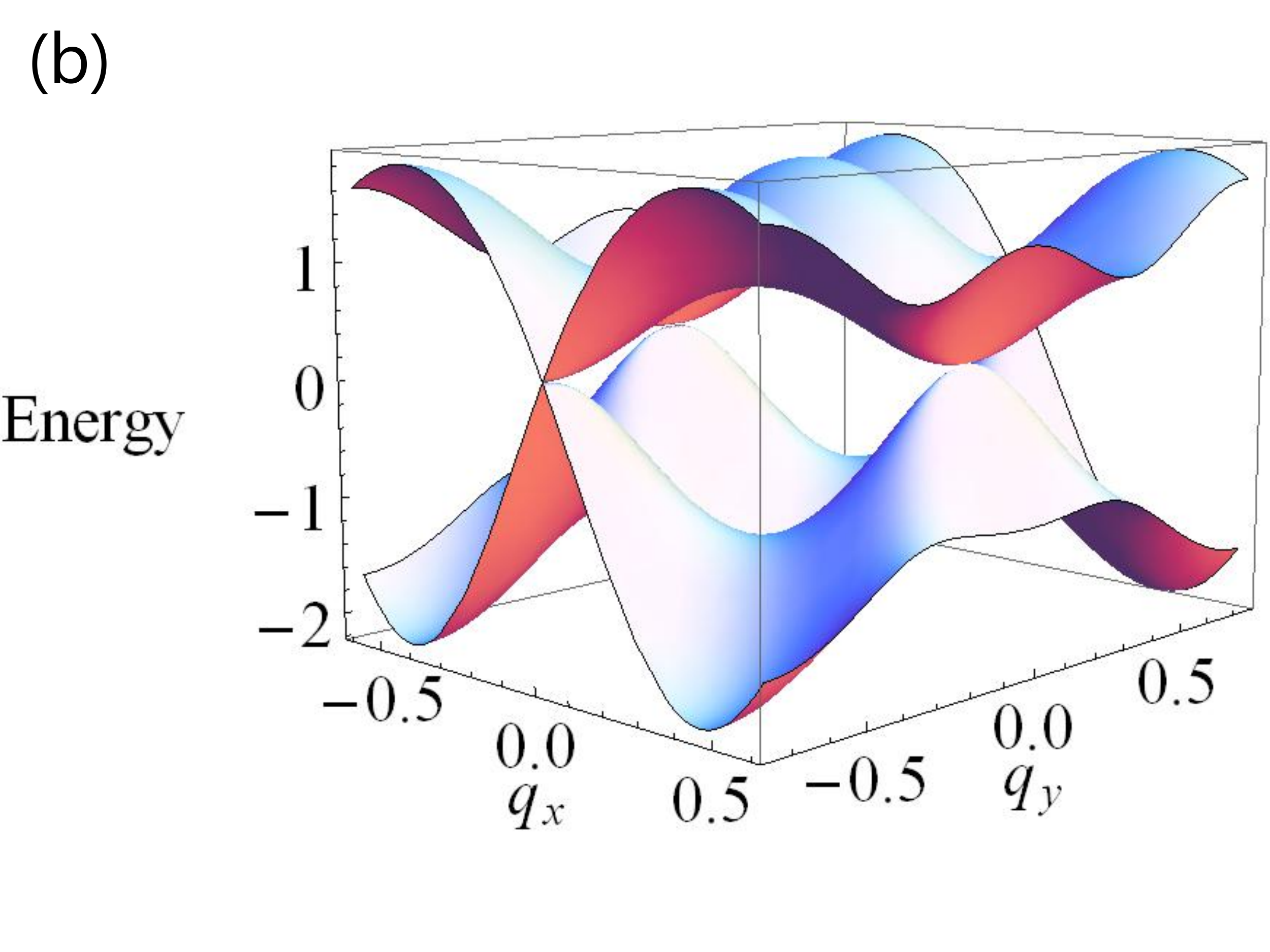}
\end{minipage}

\begin{minipage}{4cm}
	\centering
		\includegraphics[scale=0.16, angle=0]{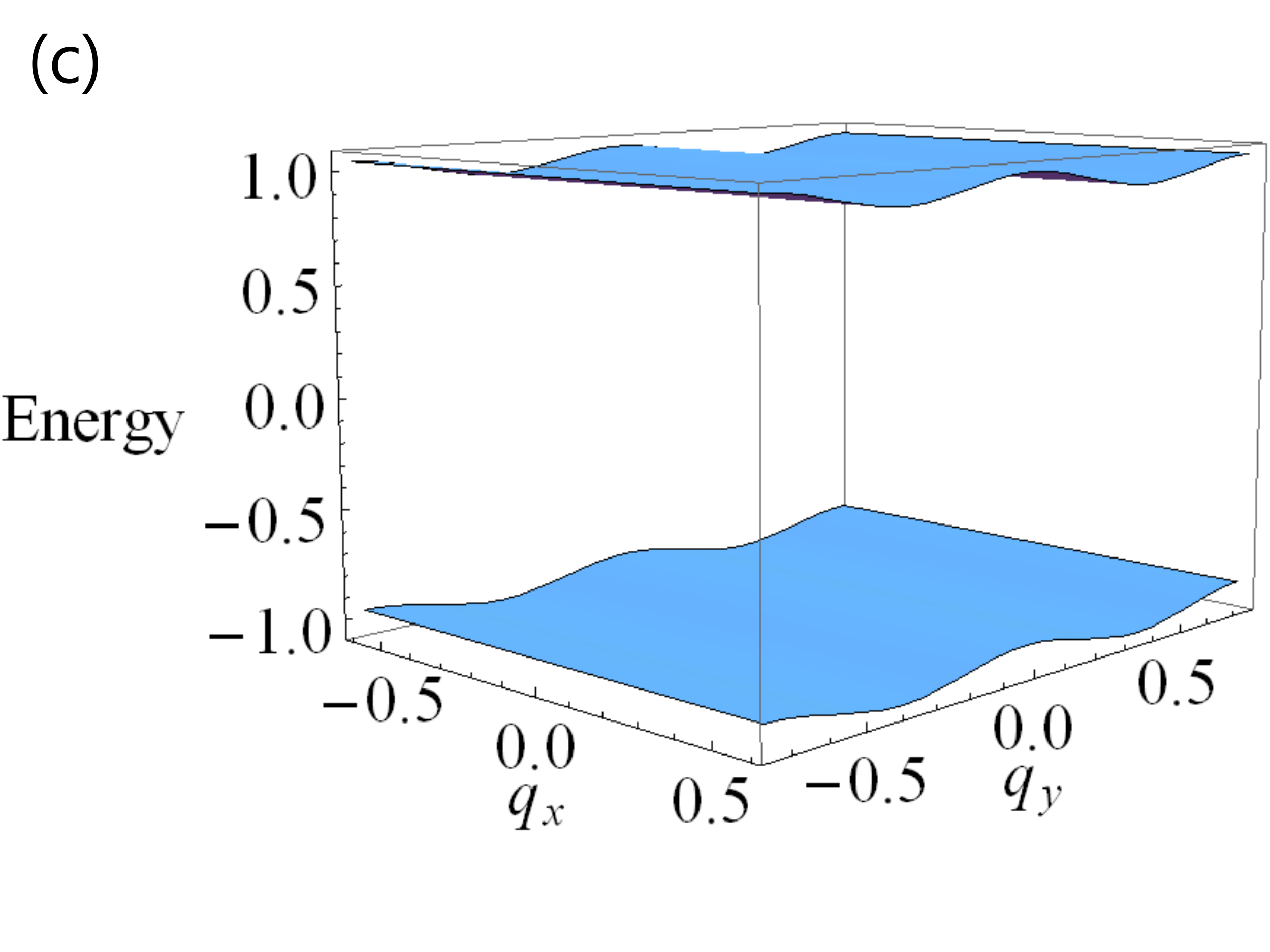}
\end{minipage}
\begin{minipage}{4cm}
	\centering
		\includegraphics[scale=0.16, angle=0]{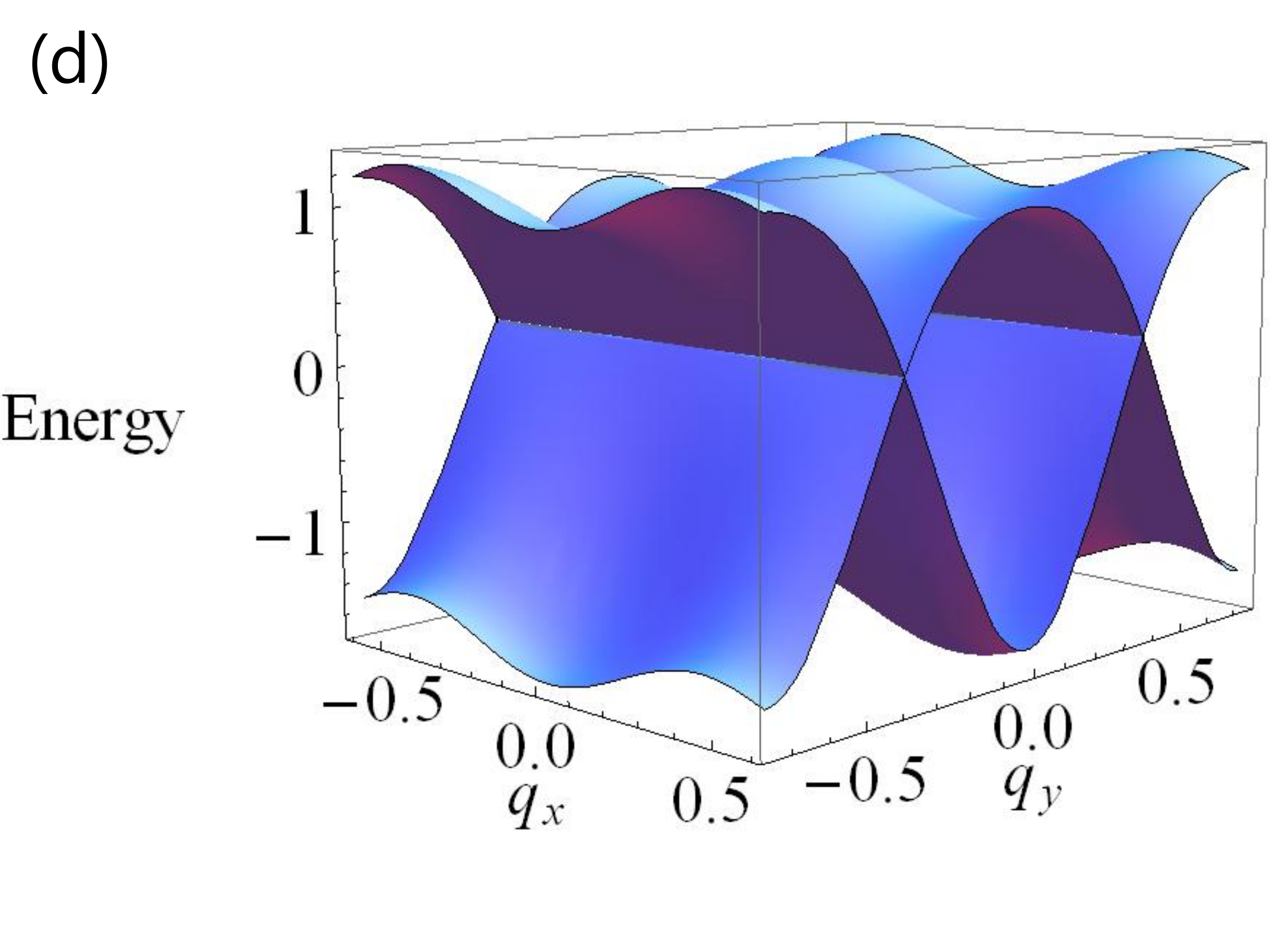}
\end{minipage}

\begin{minipage}{4cm}
	\centering
		\includegraphics[scale=0.16, angle=0]{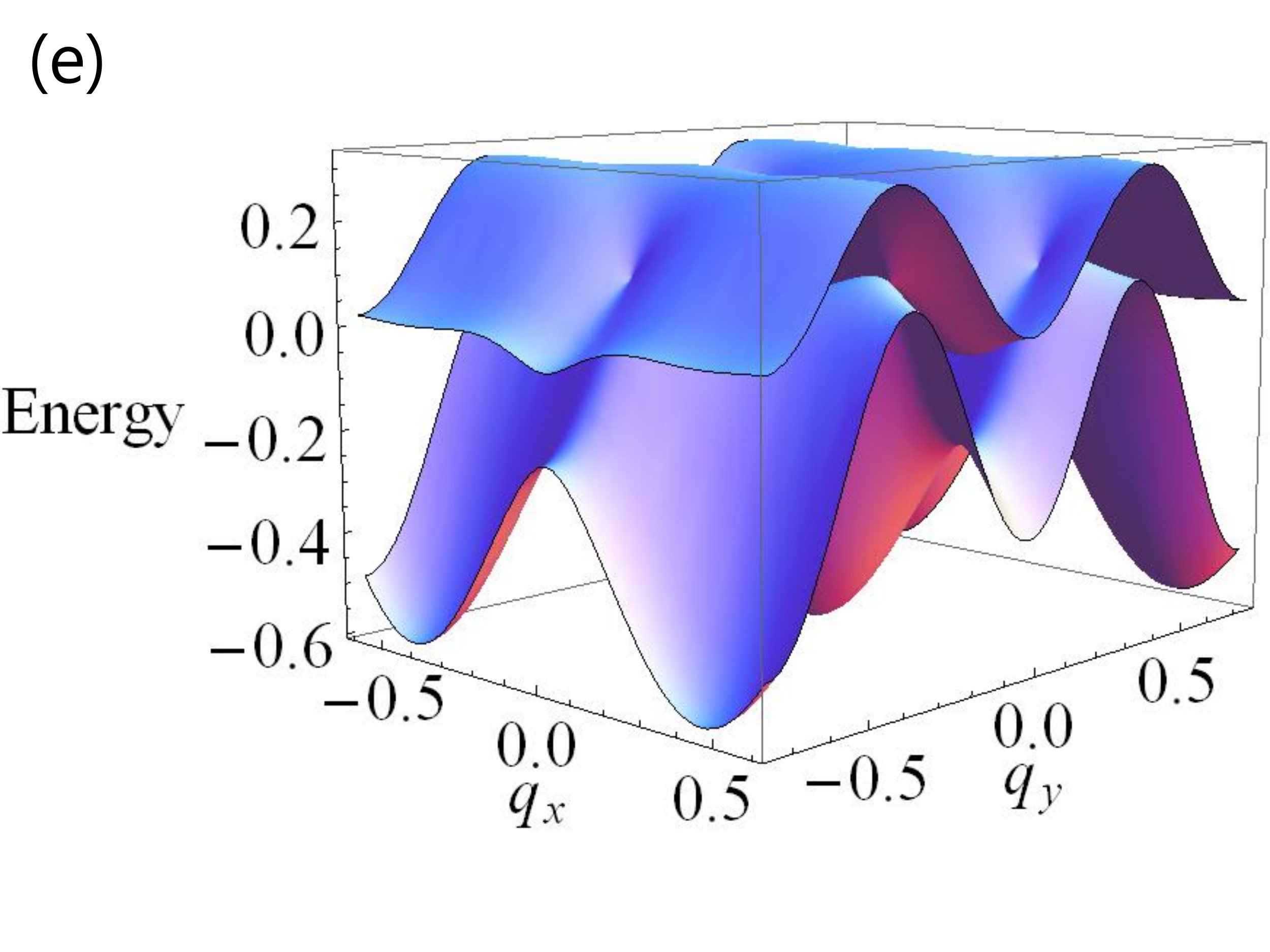}
\end{minipage}
\begin{minipage}{4cm}
	\centering
		\includegraphics[scale=0.16, angle=0]{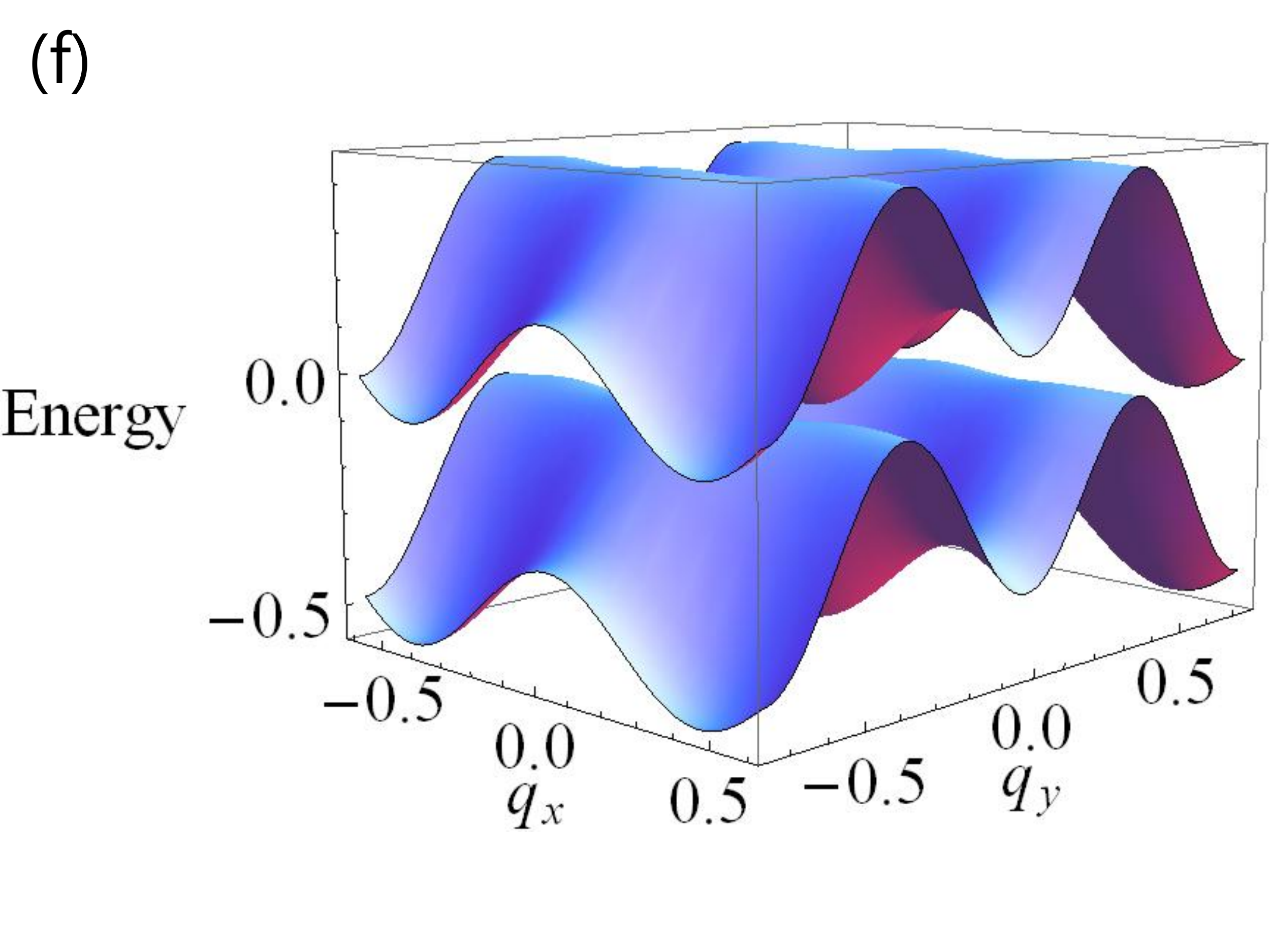}
\end{minipage}
\caption{(Color online) Energy dispersion for the shaken honeycomb optical lattice, with $k=1$, $\gamma =1$, and $\gamma'=0.1$. The labels $q_x$ and $q_y$ represent the $x$ and $y$ components of the momentum, respectively. The $x$ and $y$ axes have been chosen such that the nn vectors are given by Eq.\,(\ref{eq:nn-vec}). (a) The homogeneous case, where $\gamma_1=\gamma_{2,3}$. (b) The merged Dirac points, where $\gamma_{2,3} = \gamma_1/2$. (c) The zero dimensional case, where $\gamma_{2,3}=0$. (d) The aligned Dirac points, where $\gamma_1=0$. (e) An example of the metallic phase with $\boldsymbol{\rho} = 5.2 (\hbar / m \Omega d) \hat{e}_x$. (f) Another example of the metallic phase with $\boldsymbol{\rho} = 4.8 (\hbar / m \Omega d) \hat{e}_x$.}
\label{fig:nnn-energy-dispersion}
\end{figure}

\subsection{The signs of the hopping parameters} \label{sec02d}
\begin{figure}[h]
	\centering
		\includegraphics[scale=0.3, angle=0]{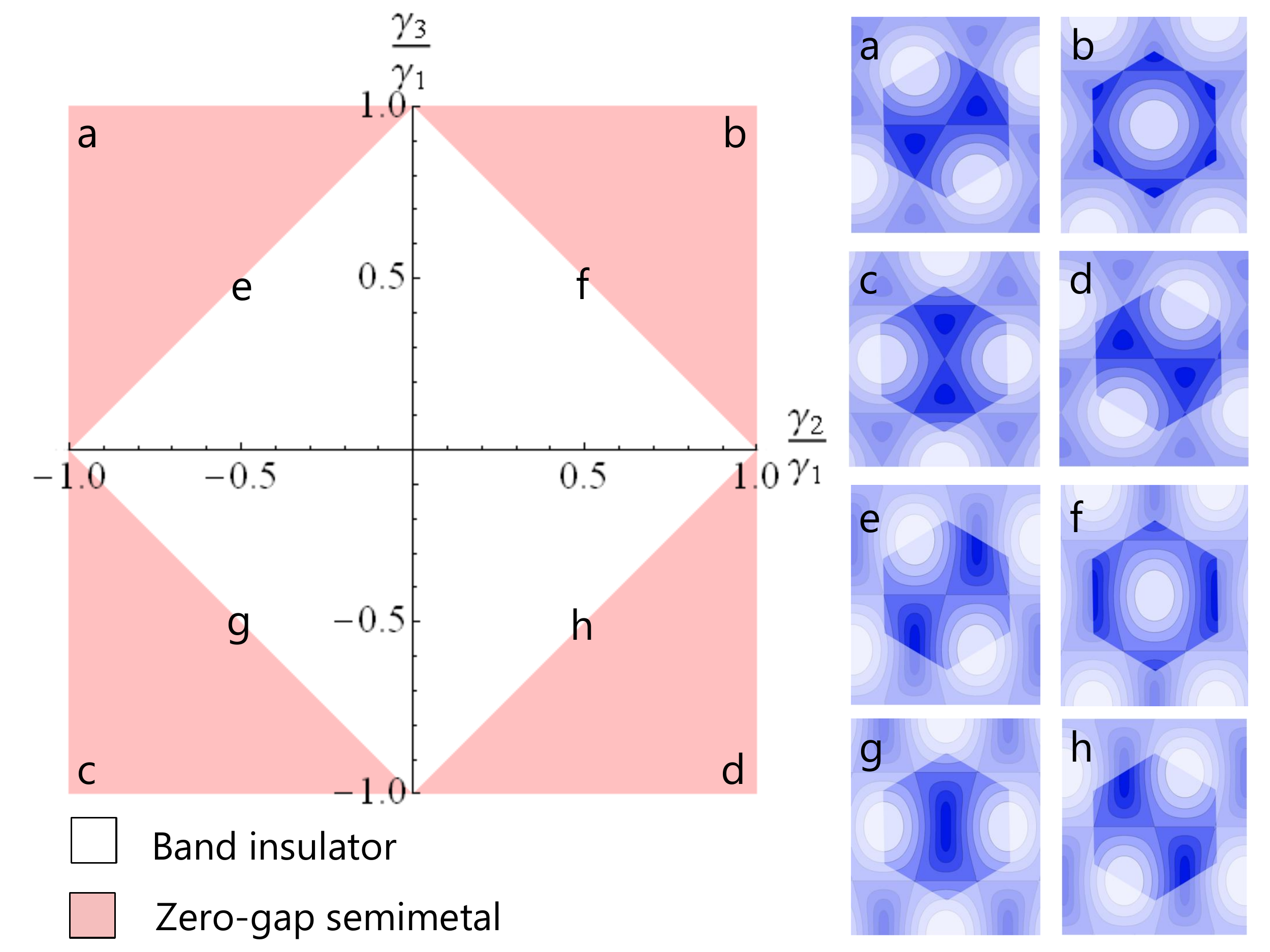}
	\caption{(Color online) Phase diagram and contour plots of the energy bands, showing the effects of the renormalized nn hopping parameters $\gamma_j$. (a) to (h) Contour plots of the energy bands, where the color scaling is arbitrary and the first Brillouin zone is the area with higher contrast. The value of the nn hopping parameters for each contour plot are given by the position of the corresponding letter in the phase diagram and  $\gamma'=0$. The dark regions indicate energies close to zero, whereas brighter regions are further away in energy from the Fermi level at half filling.}
	\label{fig:phase-diagram-sign}
\end{figure}

As already alluded to in the previous sections, the shaking of a honeycomb lattice can lead to a sign change of the hopping parameters. Quite generally, Fig.\,\ref{fig:phase-diagram-sign} shows that changing the relative signs of the nn hopping parameters results in a translation of the energy spectrum in momentum space. This effect was also mentioned in Ref. \cite{Hasegawa06}. Indeed, the relative signs determine at which of the four time-reversal invariant momenta in the first Brillouin zone the merging of Dirac points and the semimetal-insulator transition take place when $|\gamma_1|=2|\gamma_{2,3}|$. However, the sign change of the nn hopping parameters can be transformed away by a gauge transformation \cite{Hasegawa06}. Nevertheless, the sign of the nnn hopping parameter is important, since it determines whether the upper or the lower band is flattened.

\section{Interactions} \label{sec03}

Until now, we have considered single-component fermionic atoms and, due to the Pauli principle, the absence of $s$-wave interaction naturally results in an ideal Fermi lattice gas, albeit with an unusual band structure. By trapping two hyperfine states of the fermionic atoms, Hubbard-like interaction terms arise,
\begin{align}
H_{\textrm{int}}=&\sum_{\textbf{r} \in A} \sum_{\sigma, \sigma'} \frac{U}{2} \, a^\dagger_{\textbf{r},\sigma} a^\dagger_{\textbf{r},\sigma'} a_{\textbf{r},\sigma'} a_{\textbf{r},\sigma} \nonumber \\
+ &\sum_{\textbf{r} \in B} \sum_{\sigma, \sigma'} \frac{U}{2} \, b^\dagger_{\textbf{r},\sigma} b^\dagger_{\textbf{r},\sigma'} b_{\textbf{r},\sigma'} b_{\textbf{r},\sigma},
\end{align}
where the fermionic operators now acquire an additional spin index $\sigma=\{\uparrow, \downarrow\}$, which is summed over, and $U$ is the interaction energy. Naturally, in order to be able to apply the Floquet theory in the presence of interactions, the Hamiltonian $H_0$ in Eq. (\ref{eq:condition}) must now be replaced by $H = H_0 + H_{\rm int}$. This is the case in our study because we investigate the weak-coupling limit with $U\ll \gamma$. Since the interaction term commutes with the shaking, it is not renormalized, similarly to the on-site energy term proportional to $\mu$ in Eq.\,(\ref{eq:H0}). However, it has been shown that complications may arise when a multiple of the energy $U$ is in resonance with a harmonic of $\hbar\Omega$, $m\hbar\Omega=nU$, for integer $m$ and $n$. Whereas the limit \cite{Eckardt08} $m\ll n$ is not considered here because it is in contradiction with the small-$U$ large-frequency limit, critical resonances may occur for $m\gg n$ \cite{Poletti11}. Nevertheless, it has been shown in Ref. \cite{Poletti11} that these resonances, which occur in higher-order perturbation theory, are strongly suppressed in the large-$m$ limit.

In the weakly-interacting regime considered here, the ground state is adiabatically connected to that of the non-interacting system, with no broken symmetry. First, we use the Fourier transform of the creation and annihilation operators, $a_{\textbf{r},\sigma} = \mathcal{N}^{-1/2} \sum_\textbf{q} \exp(i \textbf{q} \cdot \textbf{r}) a_{\textbf{q},\sigma}$, to find the Hamiltonian in momentum space. Within a Hartree-Fock theory, we introduce a mean-field decoupling of the interaction terms,
\begin{align}
&a^\dagger_{\textbf{q}1,\sigma} a^\dagger_{\textbf{q}2,\sigma'} a_{\textbf{q}3,\sigma'} a_{\textbf{q}4,\sigma} \approx \\
\big< &a^\dagger_{\textbf{q}2,\sigma'} a_{\textbf{q}3,\sigma'} \big> a^\dagger_{\textbf{q}1,\sigma} a_{\textbf{q}4,\sigma} - \big< a^\dagger_{\textbf{q}2,\sigma'} a_{\textbf{q}4,\sigma} \big> a^\dagger_{\textbf{q}1,\sigma} a_{\textbf{q}3,\sigma'} \nonumber \\
+ \, &a^\dagger_{\textbf{q}2,\sigma'} a_{\textbf{q}3,\sigma'} \big<a^\dagger_{\textbf{q}1,\sigma} a_{\textbf{q}4,\sigma} \big> - a^\dagger_{\textbf{q}2,\sigma'} a_{\textbf{q}4,\sigma} \big< a^\dagger_{\textbf{q}1,\sigma} a_{\textbf{q}3,\sigma'} \big> \nonumber \\
- \big< &a^\dagger_{\textbf{q}2,\sigma'} a_{\textbf{q}3,\sigma'} \big> \big< a^\dagger_{\textbf{q}1,\sigma} a_{\textbf{q}4,\sigma} \big> + \big< a^\dagger_{\textbf{q}2,\sigma'} a_{\textbf{q}4,\sigma} \big> \big< a^\dagger_{\textbf{q}1,\sigma} a_{\textbf{q}3,\sigma'} \big> \nonumber,
\end{align}
such that the expectation values of both sides are equal. For the mean value we take
\begin{equation}
\langle a^\dagger_{\textbf{q},\sigma} a_{\textbf{q}', \sigma'} \rangle = \langle b^\dagger_{\textbf{q},\sigma} b_{\textbf{q}', \sigma'} \rangle = \mathcal{N} n_{\textbf{q},\sigma} \delta_{\textbf{q}, \textbf{q}'} \delta_{\sigma,\sigma'},
\label{eq:orderparameter}
\end{equation}
where $\mathcal{N}$ is the number of sites per sublattice, $n_{\textbf{q},\sigma}$ is the density of atoms with momentum $\textbf{q}$ and spin index $\sigma$, and $\delta_{\alpha,\alpha'}$ is the Kronecker delta. We then obtain the mean-field Hamiltonian
\begin{equation}
H_{MF} = H_{\textrm{eff}} - \frac{U \mathcal{N} n^2}{8} + \frac{U n}{4}  \sum_{\sigma, \textbf{q}} \big( a^\dagger_{\textbf{q},\sigma} a_{\textbf{q},\sigma} + b^\dagger_{\textbf{q},\sigma} b_{\textbf{q},\sigma} \big),
\label{eq:Ham-mf1}
\end{equation}
where the total density is defined by $n = \sum_{q,\sigma} n_{\textbf{q},\sigma} $.

The Hamiltonian (\ref{eq:Ham-mf1}) may be rewritten in a matrix form:
\begin{align}
H_{MF} = &- \frac{U \mathcal{N} n^2}{8} \label{eq:Ham-mf-matrix-ab} \\
&+ \sum_{\sigma, \textbf{q}} \left( \begin{matrix} a^\dagger_{\textbf{q},\sigma} && b^\dagger_{\textbf{q},\sigma} \end{matrix} \right)
\left( \begin{matrix} h(\mu, \textbf{q}) && f(\textbf{q}) \\ f^*(\textbf{q}) && h(\mu, \textbf{q}) \end{matrix} \right) \left( \begin{matrix} a_{\textbf{q},\sigma} \\ b_{\textbf{q},\sigma} \end{matrix} \right), \nonumber
\end{align}
where we have introduced the functions
\begin{equation}
h(\mu, \textbf{q}) = \frac{U n}{4} - \mu - \gamma' \sum^3_{i=1} \sum^3_{j=1,j\neq i} \exp[-i \textbf{q} \cdot (\textbf{d}_i - \textbf{d}_j)],
\end{equation}
and $f(\bq)$ is defined in Eq. (\ref{fq})
The Hamiltonian ($\ref{eq:Ham-mf-matrix-ab}$) can then be diagonalized by the unitary operator
\begin{equation}
\hat{{\cal U}} = \frac{1}{\sqrt{2}}\left( \begin{matrix} 1 && i f(\textbf{q})/|f(\textbf{q})| \\ f^*(\textbf{q})/|f(\textbf{q})| && -i \end{matrix} \right),
\end{equation}
which yields
\begin{widetext}
\begin{equation}
H_{MF} = - \frac{U \mathcal{N} n^2}{8} + \sum_{\sigma, \textbf{q}} \left( \begin{matrix} c^\dagger_{\textbf{q},\sigma} && d^\dagger_{\textbf{q},\sigma} \end{matrix} \right)
\left( \begin{matrix} h(\mu, \textbf{q}) - |f(\textbf{q})| && 0 \\ 0 && h(\mu, \textbf{q}) + |f(\textbf{q})| \end{matrix} \right) \left( \begin{matrix} c_{\textbf{q},\sigma} \\ d_{\textbf{q},\sigma} \end{matrix} \right).
\label{eq:Ham-mf-matrix-cd}
\end{equation}
Because the $c$ and $d$ quasiparticles are free, the partition function corresponding to the Hamiltonian ($\ref{eq:Ham-mf-matrix-cd}$) reads
\begin{equation}
Z = \exp \bigg[ \sum_{\sigma, \textbf{q}} \bigg( \log\bigg\{ 1 + \exp\big[-\beta \big(h(\mu, \textbf{q}) - |f(\textbf{q})| \big) \big] \bigg\} + \log\bigg\{ 1 + \exp\big[-\beta \big(h(\mu, \textbf{q}) + |f(\textbf{q})| \big) \big] \bigg\} \bigg) \bigg],
\end{equation}
where $\beta = (k_B T)^{-1}$ with $k_B$ denoting the Boltzmann's constant and $T$ the temperature.

The total number of particles $N$ is given by $(1/\beta) \partial \log Z / \partial \mu$, and one obtains
\begin{equation}
N = \sum_{\sigma,\textbf{q}} \bigg( \frac{1}{1 + \exp \big[\beta \big(h(\mu, \textbf{q}) - |f(\textbf{q})| \big) \big]} + \frac{1}{1 + \exp\big[\beta \big(h(\mu, \textbf{q}) + |f(\textbf{q})| \big) \big]} \bigg).
\label{eq:nrparticles}
\end{equation}
Since the expression inside the sum does not depend on spin, summing over $\sigma$ yields a factor 2.
One recognizes in Eq.\,(\ref{eq:nrparticles}) the Fermi-Dirac distribution function $N_{FD}(x)=[1+\exp(x)]^{-1}$.
The number of particles $N$ is related to the density $n$, which is defined here as the number of particles per lattice site, i.e. $n = N / 2\mathcal{N}$. Converting the sum over $\textbf{q}$ into an integral, 
the following self-consistent equation for the density is derived
\begin{equation}
n(\mu) = \frac{1}{V_{1BZ}} \int\limits_{1BZ} d^2\textbf{q} \bigg\{N_{FD}\big[\beta \big(h(\mu, \textbf{q}) - |f(\textbf{q})| \big) \big] + N_{FD}\big[\beta \big(h(\mu, \textbf{q}) + |f(\textbf{q})| \big) \big] \bigg\},
\label{eq:density-interaction-os}
\end{equation}
where the integral is restricted to the first Brillouin zone, the surface of which is $V_{1BZ}$.
\\
\end{widetext}

\begin{figure}[h]
\begin{minipage}{4cm}
	\centering
		\includegraphics[scale=0.16, angle=0]{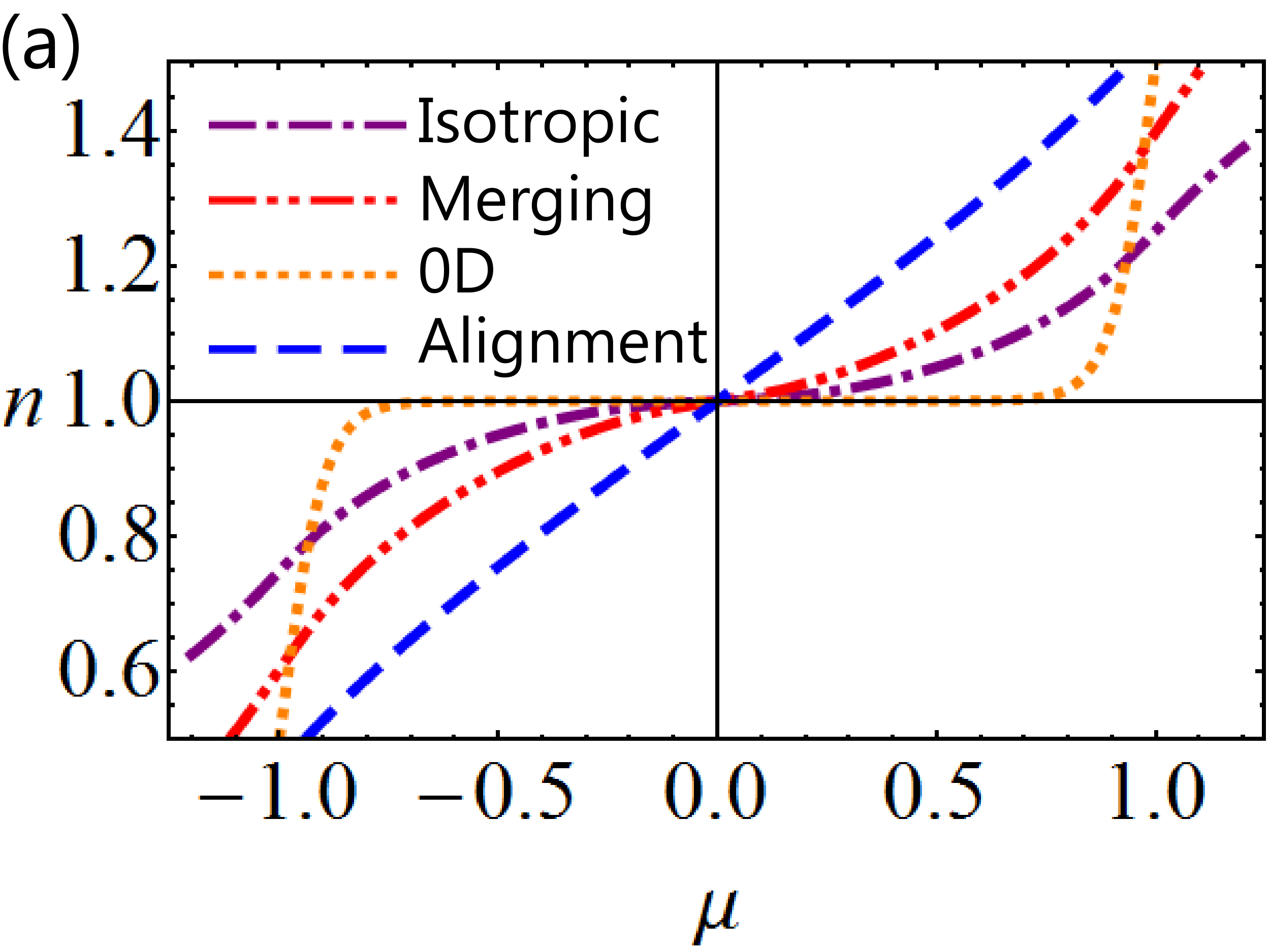}
\end{minipage}
\begin{minipage}{4cm}
	\centering
		\includegraphics[scale=0.16, angle=0]{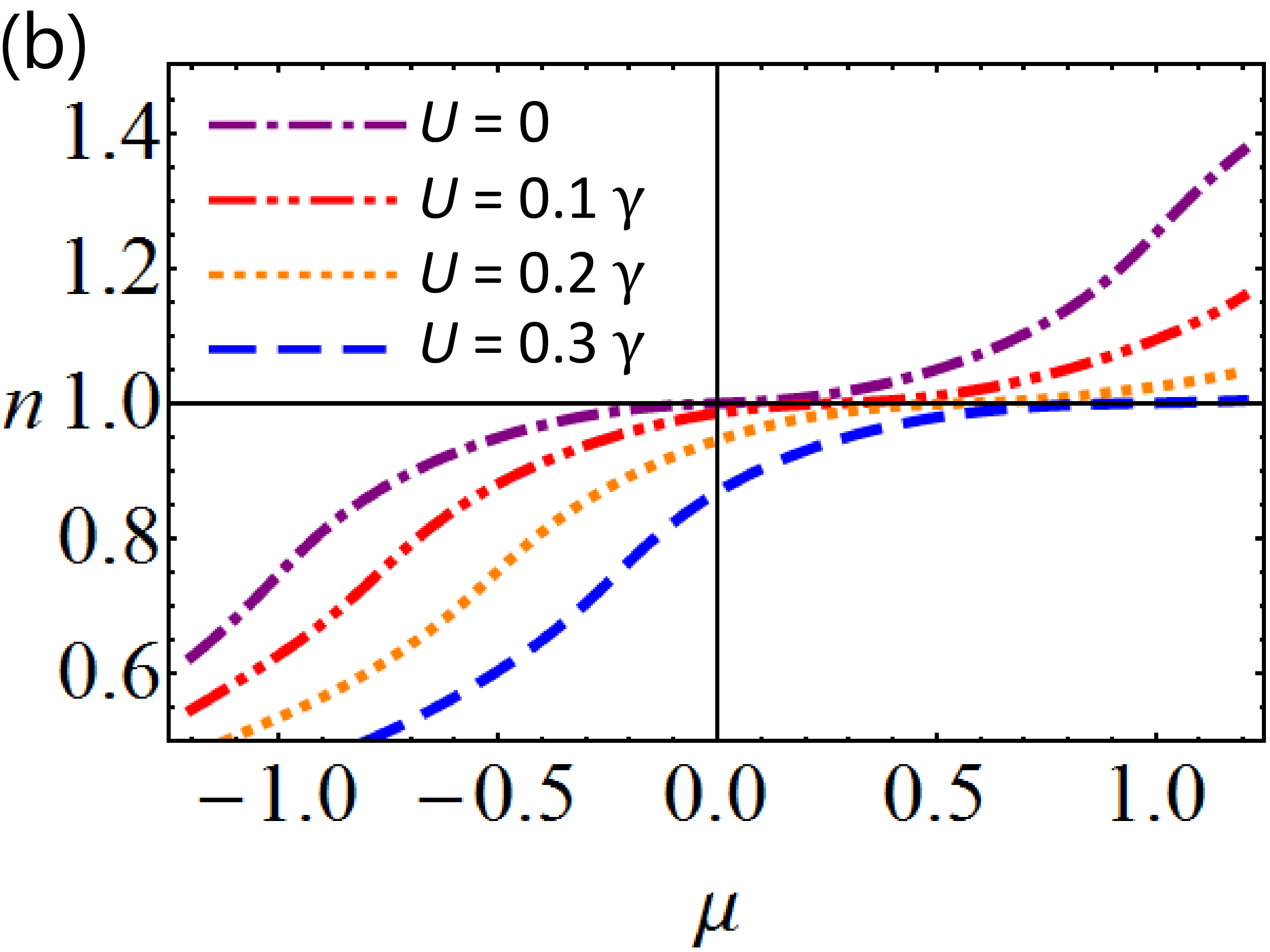}
\end{minipage}

\begin{minipage}{4cm}
	\centering
		\includegraphics[scale=0.16, angle=0]{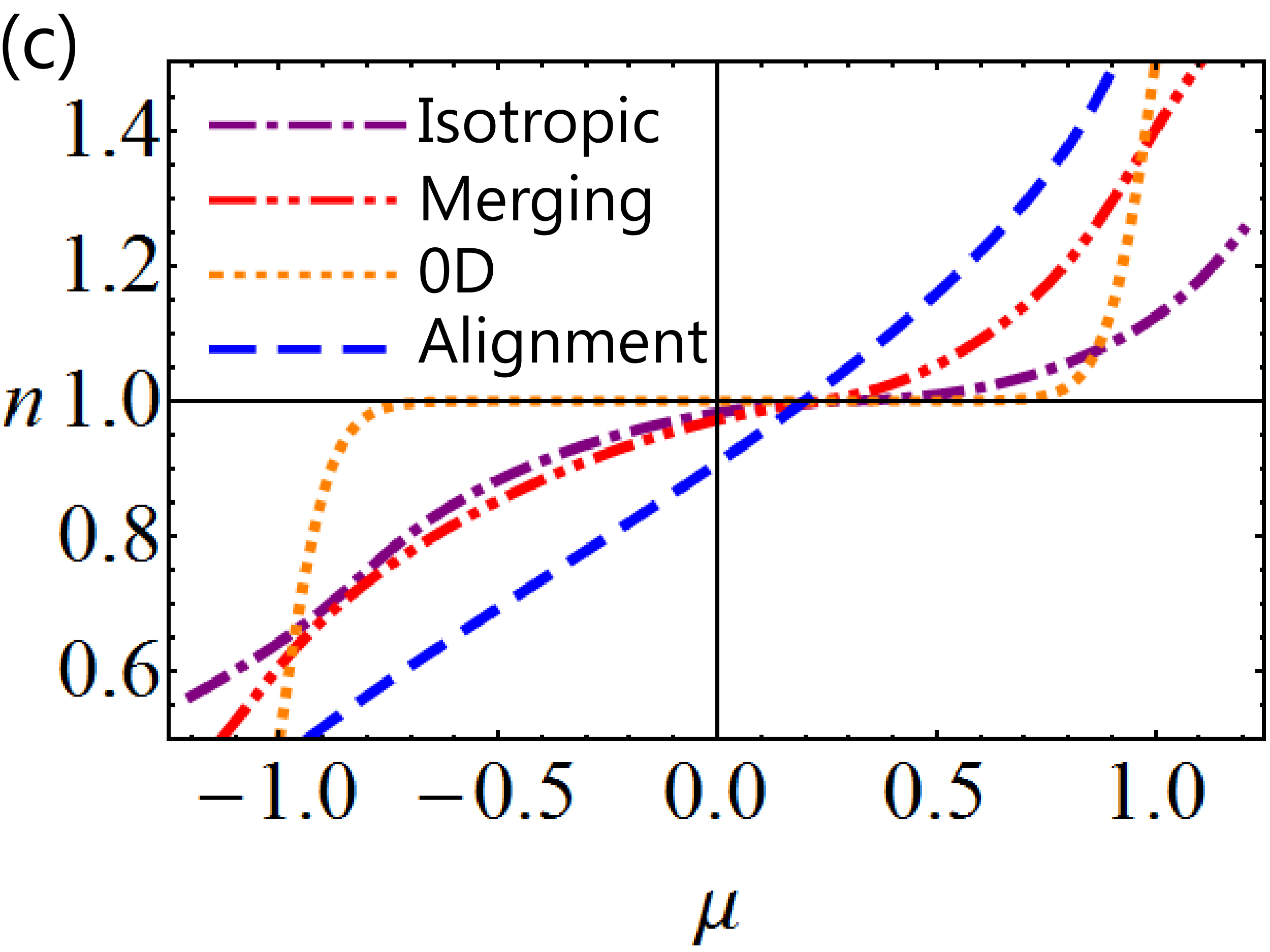}
\end{minipage}
\begin{minipage}{4cm}
	\centering
		\includegraphics[scale=0.16, angle=0]{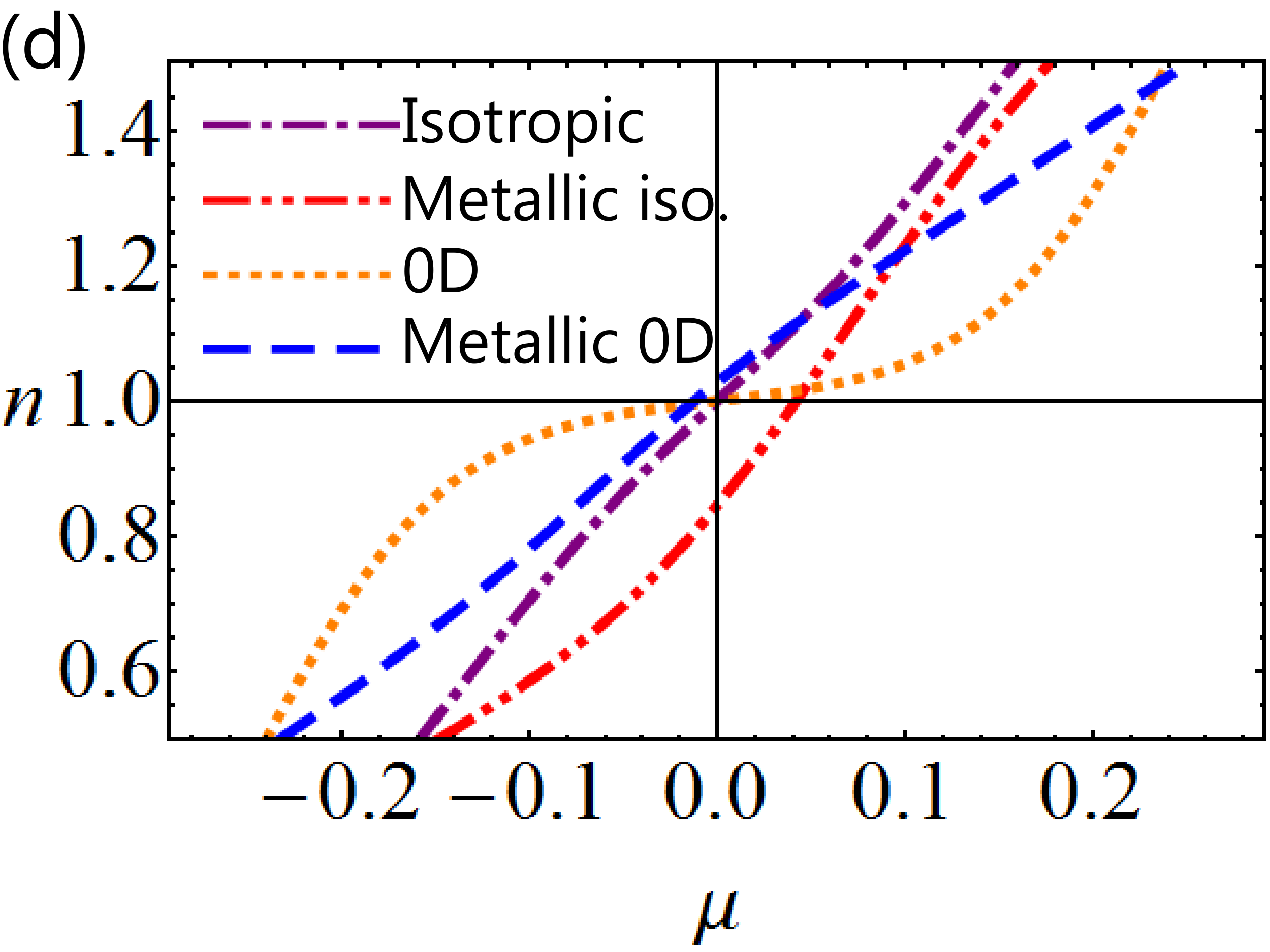}
\end{minipage}
\caption{(Color online) Density $n$ as a function of the chemical potential $\mu$. Unless specified otherwise in the figure, the nn hopping parameters $\gamma_{2,3}=\gamma_1=\gamma =1 $, the nnn hopping parameter $\gamma'=0$, the interaction strength $U=0$, and the inverse temperature $\beta=20$. (a) Effect of the renormalization of the nn hopping parameters. (b) Effect of the interaction strength $U$ for the isotropic case. (c) Effect of the nnn hopping parameter $\gamma'=0.1$ in the shaken lattice. (d) The metallic phase. For the isotropic cases, $\boldsymbol{\rho} = 5.2 (\hbar / m \Omega d) \hat{e}_x$, whereas for the 0D cases $\boldsymbol{\rho} = 4.8 (\hbar / m \Omega d) \hat{e}_x$. These systems are in the metallic phase for $\gamma'=0.1$, whereas for $\gamma'=0$ they are in the semi-metallic and the insulating phase, respectively.}
\label{fig:density-mu}
\end{figure}

In Fig.\,\ref{fig:density-mu}(a), the density $n(\mu)$ is plotted for several values of $\gamma_{2,3}/\gamma_1$. For the isotropic case, $\gamma_{2,3}/\gamma_1 = 1$, the result of Zhu \textit{et al}. is reproduced \cite{Zhu07}. For the 0D limit, the flat line due to the gap in the spectrum is clearly visible at the chosen temperature. Fig.\,\ref{fig:density-mu}(b) confirms that repulsive interactions lead to a lower density than in a system without interactions for the same chemical potential. Fig.\,\ref{fig:density-mu}(c) agrees with the observation that the nnn hopping breaks the particle-hole symmetry. This effect is also visible in Fig.\,\ref{fig:density-mu}(d), where the dependence of the density on the chemical potential is calculated for a shaking vector where the system is in the zero-gapped semimetallic phase for $\gamma'=0$ and in the metallic phase for $\gamma'=0.1$.

\section{Possibilities for experimental observation} \label{sec04}

Honeycomb optical lattices have recently been realized experimentally, although the existing setups have only been used to investigate bosonic atoms \cite{Honeycomb10, Soltan11}.
Shaking of a lattice has been experimentally implemented in a one-dimensional one by a periodic modulation on the position of the reflecting mirrors \cite{Zenesini09}. For a honeycomb lattice, the shaking could be realized by means of an acousto-optical device, as proposed for a triangular lattice in Ref. \cite{Eckardt10}.

The magnitude of the nn hopping parameter $\gamma$ in a honeycomb optical lattice has been evaluated in Ref. \cite{Lee09},
\begin{equation}
\gamma \approx 1.861  E_R \left(\frac{V_0}{E_R}\right)^{3/4} \exp\left( -1.582 \sqrt{\frac{V_0}{E_R}} \right) ,
\label{eq:size-gamma}
\end{equation}
in terms of the recoil energy $E_R=\hbar^2 k^2 / 2 m$ and the magnitude of the potential barrier between nearest-neighbor lattice sites $V_0$. The magnitude of the nnn hopping parameter $\gamma'$ is not yet known, but could be determined from numerical band structure calculations. In a typical experimental situation, we expect the ratio $\gamma'/\gamma$ to be in the $5-10\%$ range, in agreement with the parameter chosen in the discussion of Sec.\,\ref{sec02c}. 

\begin{figure}[h]
\centering
\includegraphics[scale=0.3, angle=0]{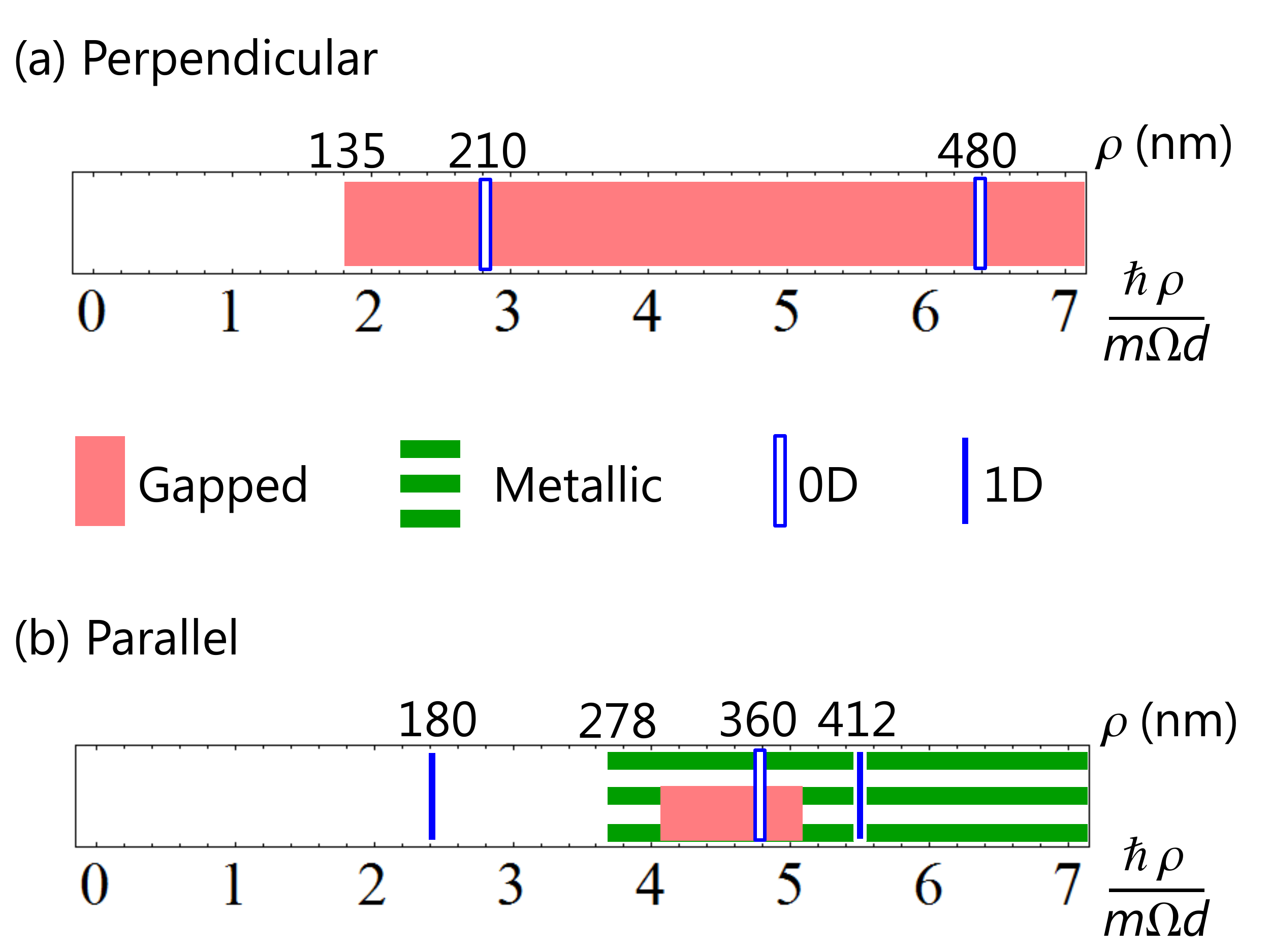}
\caption{(Color online) Overview of the different phases as a function of the shaking amplitude. The bottom scale gives the size of the argument of the Bessel function, whereas the values for $\rho$ in nm correspond to the optical lattice discussed in Sec.\,\ref{sec02c} with $\gamma'= 0.1\gamma$, $\Omega/2\pi=6$kHz, $k=830$nm, and containing $^{40}$K atoms. (a) Shaking perpendicular to one of the nn hopping directions. (b) Shaking parallel to one of the nn hopping directions.}
\label{fig:phases-linear}
\end{figure}

In a typical experiment, the shaking amplitude would be increased from zero to a finite value. Fig.\,\ref{fig:phases-linear} shows in which order the system goes through the different phases and dimensions upon increasing the shaking amplitude. Here, also the values of the shaking amplitude required for the dimensional crossovers are given for the same system as discussed in Sec.\,\ref{sec02c} and $\gamma'=0.1\gamma$. If the shaking direction is perpendicular to one of the nn vectors, the system will be in the gapped insulating phase beyond a certain value of the shaking amplitude, since the Bessel function crosses the value 0.5 only once and never obtains the value -0.5.  If the shaking is parallel to one of the nn vectors, the system will be in the metallic phase beyond a certain value of the shaking amplitude. Nevertheless, it is still possible to induce a merging of Dirac points and to open up a gap in the spectrum, since the nn hopping parameters are renormalized such that the value of one of them will in general differ from that of the other two. However, whether the system is actually driven into an insulating phase or remains metallic depends on the precise value of the ratio $\gamma'/\gamma$.

In experiments, an overall harmonic trapping potential is imposed to confine the atoms. It is described by
\begin{equation}
V_{\textrm{trap}}(\textbf{r}) = \frac{1}{2} m \omega_\textrm{trap}^2 \textbf{r}^2 ,
\end{equation}
where $\omega_\textrm{trap}$ is the trapping frequency, and $\textbf{r}$ is the position measured from the center of the trap. By applying the local density approximation (LDA), one finds that the chemical potential evolves radially according to $\mu \rightarrow \mu - V_{\textrm{trap}}(\textbf{r}^2)$.

\begin{figure}[t]

\begin{minipage}{4cm}
	\centering
		\includegraphics[scale=0.16, angle=0]{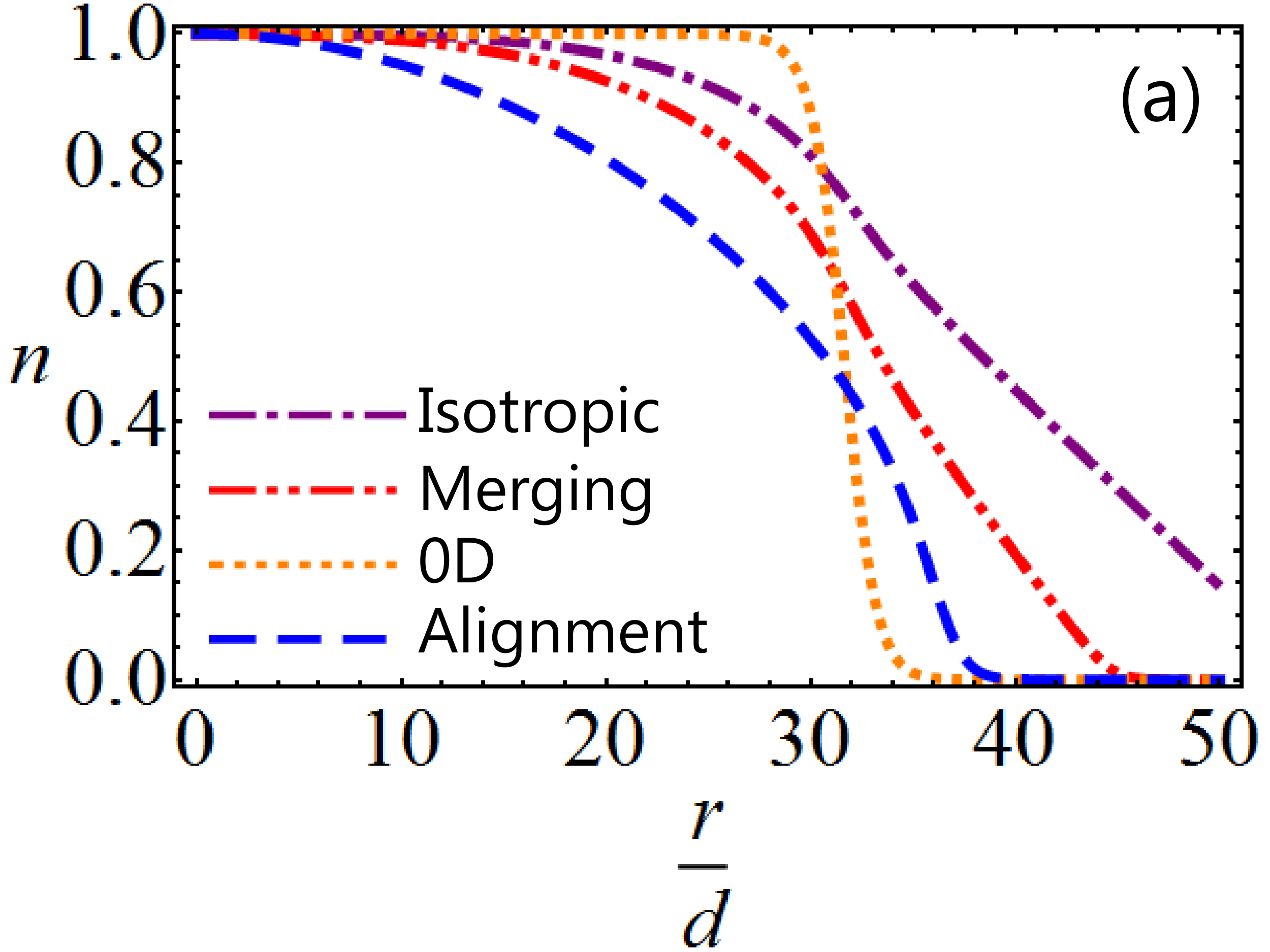}
\end{minipage}
\begin{minipage}{4cm}
	\centering
		\includegraphics[scale=0.16, angle=0]{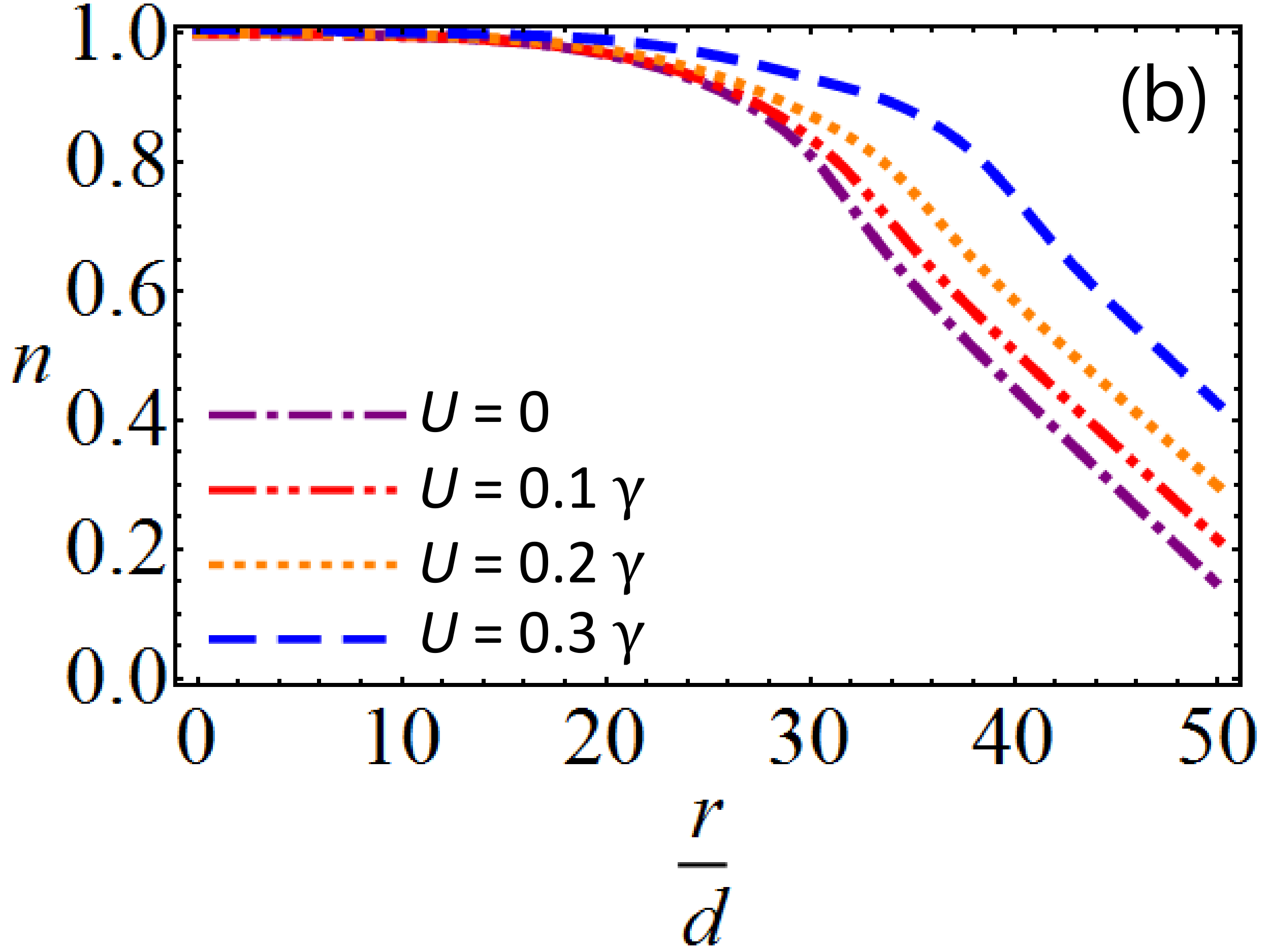}
\end{minipage}

\begin{minipage}{4cm}
	\centering
		\includegraphics[scale=0.16, angle=0]{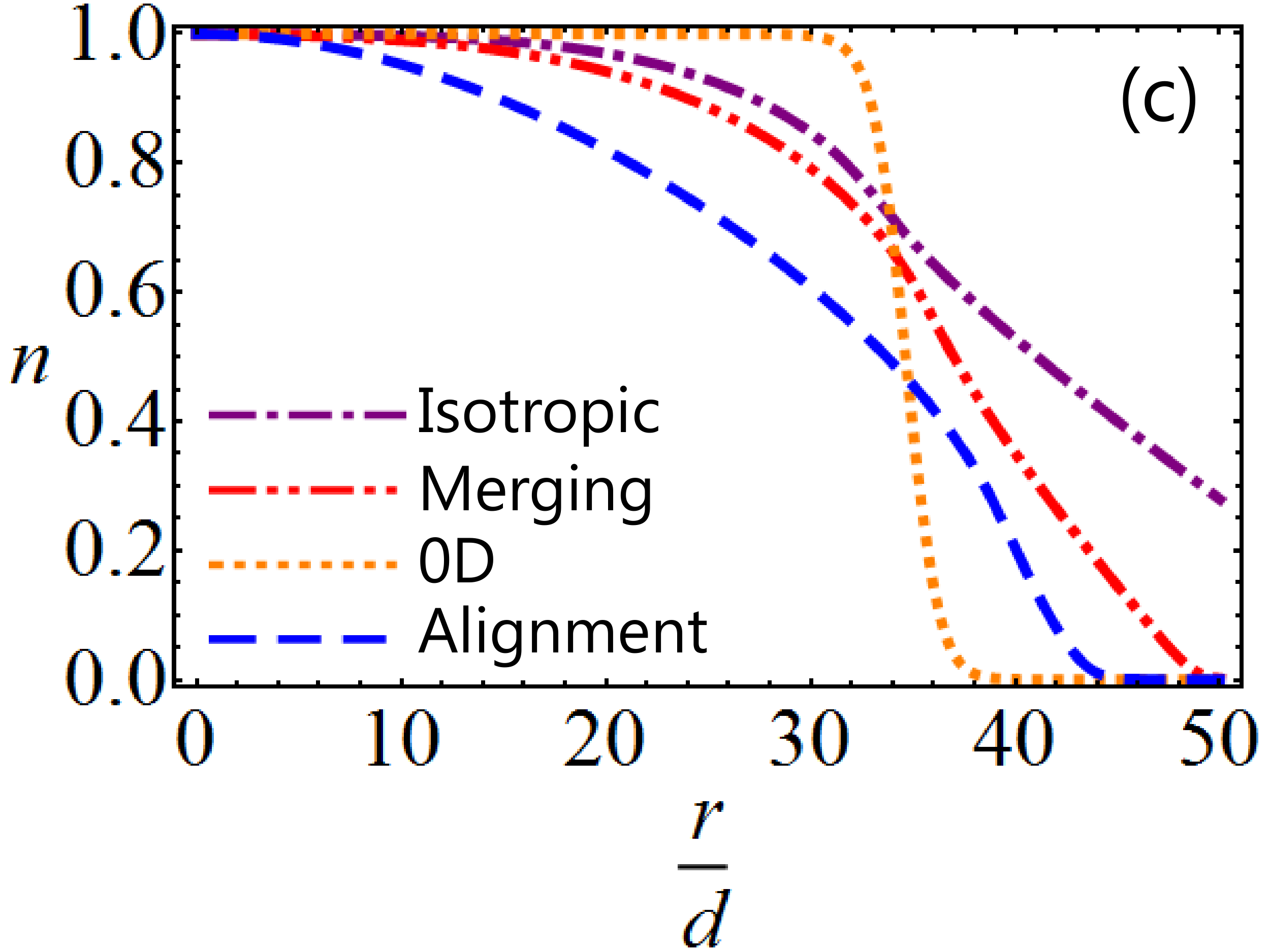}
\end{minipage}
\begin{minipage}{4cm}
	\centering
		\includegraphics[scale=0.16, angle=0]{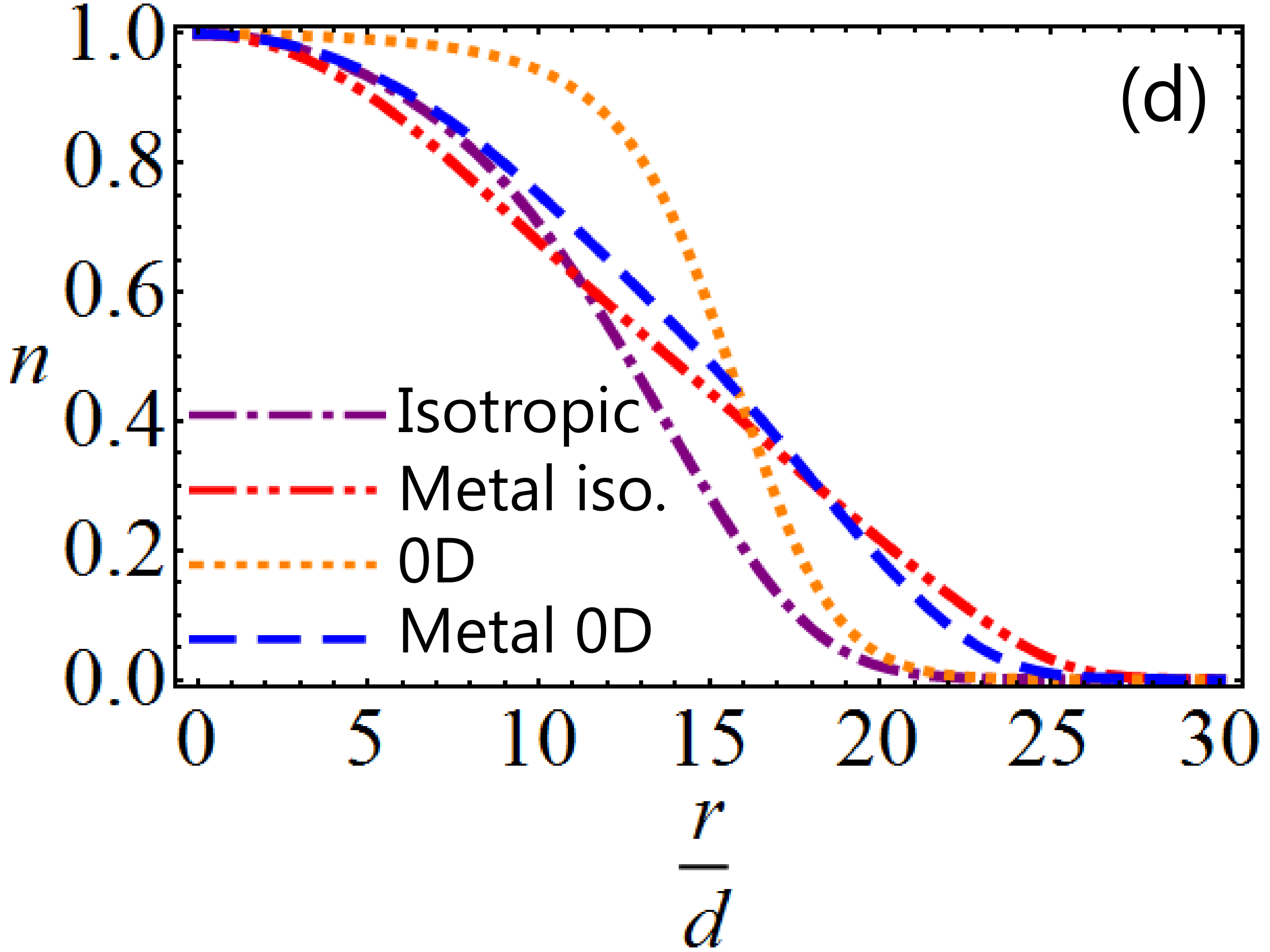}
\end{minipage}
\caption{(Color online) Density $n$ as a function of the distance from the trap's centre $r=|\textbf{r}|$, which is expressed in units of the nearest-neighbor distance $d$. The trapping frequency has been chosen such that the trapping potential is given by $V_{\textrm{trap}}(\textbf{r}) = 0.001 \gamma \textbf{r}^2 / d^2$. The chemical potential $\mu$ for each case has been chosen such that the density at the trap's centre $n$ is one particle per site. This corresponds to half-filling, since we consider 2 species of fermions. Unless specified otherwise in the figure, the nn hopping parameters $\gamma_{2,3}=\gamma_1=\gamma =1 $, the nnn hopping parameter $\gamma'=0$, the interaction strength $U=0$, and the inverse temperature $\beta=20$. (a) Effect of the renormalization of the nn hopping parameters. (b) Effect of the interaction strength $U$ for the isotropic case. (c) Effect of the nnn hopping parameter $\gamma'=0.1$ in the shaken lattice. (d) The metallic phase. For the isotropic cases, $\boldsymbol{\rho} = 5.2 (\hbar / m \Omega d) \hat{e}_x$, whereas for the 0D cases $\boldsymbol{\rho} = 4.8 (\hbar / m \Omega d) \hat{e}_x$. These systems are in the metallic phase for $\gamma'=0.1$, whereas for $\gamma'=0$ they are in the semi-metallic and the insulating phase, respectively.}
\label{fig:density-r}
\end{figure}

Fig.\,\ref{fig:density-r}(a) shows the density profile for several ratios of $\gamma_{2,3}/\gamma_1$, without nnn hopping or interactions. The case with $\gamma_{2,3}/\gamma_1=0$, when the system is in the extreme limit of the band insulating phase, can be well distinguished from the other cases. Fig.\,\ref{fig:density-r}(b) shows that stronger interactions lead to a higher density away from the center of the trap. This effect becomes visible when the density starts to deviate from one particle per lattice site. Next-nearest-neighbor hopping leads to a higher density at the edge of the cloud compared to the case without nnn hopping, which can be seen from comparing Figs.\,\ref{fig:density-r}(a) and (c) and from Fig.\,\ref{fig:density-r}(d). The latter shows the effect of nnn hopping on the density profile for the case where the nnn hopping gives rise to the metallic phase for two different shaking vectors. In the first case, $\boldsymbol{\rho} = 5.2 (\hbar / m \Omega d) \hat{e}_x$, which gives $\gamma_1 \approx \gamma_{2,3}$, such that without nnn hopping, the system is in the zero-gapped semi-metallic phase and the Dirac points are located very close to the corners of the first Brillouin zone. In the second case, $\boldsymbol{\rho} = 4.8 (\hbar / m \Omega d) \hat{e}_x$, which results in $\gamma_{2,3} \approx 0$, such that without nnn hopping the system is in the insulating phase and the two energy bands are almost flat.

We emphasize that, in the present paper, we only discuss weak correlations that adiabatically affect the density. However, when increasing further the onsite interaction, one may expect correlated phases with inhomogeneous density, even at half-filling. 
A detailed study of these correlated phases is a vast research issue that is yet ongoing and that is beyond the scope of 
the present paper. Here, we only provide a glimpse on how the density, which we discussed above in the weak-coupling limit, may evolve 
in view of some phases studied in the literature.
Mean-field calculations indicate a transition to an antiferromagnetic state above a value of $U/\gamma\simeq 2.2$ \cite{Sorella92}, whereas more sophisticated quantum Monte-Carlo calculations indicate an intermediate spin-liquid phase between the semimetal and
the anti-ferromagnetic phase \cite{Meng10}. The spin-liquid phase may be viewed as a Mott insulator with a charge localization on the lattice sites, and recent slave-rotor calculations indicate that such spin-liquid phases dominate the phase diagram for $\gamma_1>\gamma_{2,3}$ \cite{Wang11}, which is the parameter range where the Dirac points would merge in the absence of interactions. The precise transition between the weakly-interacting liquid phases and these strongly-correlated Mott insulators could in principle
be determined with the help of the above-mentioned density measurements.

A more promising technique for detecting Dirac-point motion is momentum-resolved Raman spectroscopy. This technique has been proposed as an equivalent of angle-resolved photoemission spectroscopy for cold atom systems \cite{Dao07}. It has not yet been realized experimentally, while momentum-resolved radio-frequency spectroscopy, which is a very similar technique, has already been implemented \cite{Stewart08}. Notice further that another very similar technique, momentum-resolved Bragg spectroscopy, has been applied to ultracold bosonic atoms in a static optical lattice by Ernst \textit{et al}. \cite{Ernst10}.
Momentum-resolved spectroscopy can allow us to indirectly visualize the band structure. In momentum-resolved Raman spectroscopy, two laser pulses with frequencies $\omega_1$ and $\omega_2$ are irradiated upon the system. If the frequency difference is in resonance with
a transition $\omega_{hf}$ between atomic hyperfine states,
$\omega_1-\omega_2=\omega_{hf}$, some atoms are excited in a second-order process to the higher hyperfine state. 
Then, with state-selective time-of-flight measurements, the dispersion of the atoms in the new state are measured, from which the dispersion of the original atoms can be derived. When the atoms are confined in a trapping potential and the laser pulses are focussed on the center of the trap, the quality of the results obtained by Raman spectroscopy is comparable to those of a homogeneous system \cite{Dao07}. Furthermore, Raman spectroscopy yields better results for a system with strong interactions compared to standard time-of-flight measurements \cite{Dao07}. 

Notice that momentum resolved Raman spectroscopy was originally proposed to be applied to a gas of ultracold fermionic atoms
at equilibrium and not for a shaken lattice. We therefore discuss, in this final paragraph, why we think that this technique may 
also be applied to the present case. Naturally, as long as the frequencies of the additional lasers in the Raman-spectroscopy setup
are small with respect to the shaking frequency, $\omega_1,\omega_2\ll \Omega$, even the full system satisfies the condition 
(\ref{eq:condition}) for the validity of Floquet theory. As in the case of interactions, one needs, however, to avoid resonances between
the different laser frequencies that could become critical \cite{Poletti11}. The opposite limit, in which the laser frequencies and that
of the hyperfine transition are larger than the shaking frequency, is more delicate. However, even then, the shaken system remains
at quasi-equilibrium as long as the intensities of the lasers used in Raman spectroscopy are weak, such that they only constitute a
small perturbation. The atomic dynamics probed even at high frequencies is therefore still that of the atoms at quasi-equilibrium, with
the band strucure obtained from Floquet theory.
Furthermore, in the experimental studies by Zenesini \textit{et al}. \cite{Zenesini09}, time-of-flight measurements were used to determine the momentum distribution of bosonic atoms in a shaken lattice. 
Apart from the time-scale considerations, there are also some length scales that need to be taken into account. There are indeed
two requirements for the correct size of the focus of the laser beams. On the one hand, it needs to be larger than the lattice spacing, such that sufficiently many atoms can be excited, while on the other hand the focus of the beams should be small enough, in order to have an approximately flat trapping potential inside the focus area. In addition, choosing the length of the pulses could possibly be a problem, since for shorter pulses, the excited atoms will be less affected by the lattice potential, whereas for longer pulses more atoms can be excited, leading to a stronger signal. 

\section{Conclusions} \label{conc}

In conclusion, we have investigated the band engineering of fermionic atoms in an optical honeycomb lattice with the help of a periodic shaking of the lattice. If the shaking frequency $\Omega$ is large enough, i.e. if $\hbar\Omega$ constitutes the largest energy scale in the system, the Floquet theory may be applied and the system is at quasi-equilibrium in the sense that the atoms cannot follow the rapid motion associated with the shaking. Depending on the direction of the shaking, one may render the hopping amplitudes in the quasi-static lattice anisotropic, due to a renormalization of the nn and nnn hopping parameters by Bessel functions that go through zero and change sign. As a consequence, dimensional crossovers can be induced in absence of the nnn hopping. For a shaking direction parallel to
one of the nn vectors (such as e.g. $\textbf{d}_1$), one can make one of the nn hopping parameters vanish, $\gamma_1 \rightarrow 0$.
The system then undergoes a transition from 2D to 1D, while the Dirac points align simultaneously. 
Shaking in the perpendicular direction ($\perp \textbf{d}_1$) allows one to decrease two nn hopping amplitudes simultaneously
while maintaining $\gamma_1$ unrenormalized. In this case, 
a dimensional crossover from 2D to 0D is induced for $\gamma_{2,3} \rightarrow 0$, leading to two flat energy bands, beyond the merging of Dirac points \cite{Hasegawa06,Montambaux09}, which occurs at $|\gamma_1| = 2 |\gamma_{2,3}|$. A nonzero value of $\gamma'$ breaks the particle-hole symmetry and leads to a coupling among the 1D chains and the 0D dimers, for the $\gamma_1 = 0$ and $\gamma_{2,3} = 0$ cases, respectively, and thus to a weak 2D dispersion. The merging and the alignment of Dirac points, however, are not affected. 
Moreover, for a shaking direction parallel to $\textbf{d}_1$, one pair of nnn hopping amplitudes [$\pm(\textbf{d}_2-\textbf{d}_3)$]
remains unrenormalized, and its relative importance is thus enhanced when compared to the decreasing nn hopping amplitudes. In
this limit, beyond the semi-metallic and the band-insulating phases, a novel metallic phase can appear that consists of particle and
hole pockets with a non-vanishing density of states even at half-filling. 

Furthermore, we have investigated the role of weak repulsive on-site interactions. The resulting ground state is then adiabatically connected to that of the non-interacting system, and we have self-consistently calculated the dependence of the atomic density on the (local) chemical potential. The density profiles of the different phases, e.g. the gapless semimetal or the gapped band insulator, and the different dimensionality may be measured experimentally by in-situ density measurements. Moreover, momentum-resolved Raman spectroscopy may be a promising technique to measure the band structure associated with these different phases.

\section*{Acknowledgements}

We thank Gilles Montambaux, Guangquan Wang, Andreas Hemmerich, and Marco Di Liberto for fruitful discussions. We also thank Christoph \"{O}lschl\"{a}ger for informing us about an error in the calculations. This work was financially supported by the ANR project NANOSIM GRAPHENE under Grant No. ANR-09-NANO-016 and by the Netherlands Organization for Scientific Research (NWO).

\appendix
\section{Effective Hamiltonian} \label{app1}

For the studied case, $H(t) - \hbar \partial_t F(t) = H_0$, where $H_0$ was given in Eq.\,(\ref{eq:H0}). Since the nn hopping is usually larger than the nnn hopping, $\gamma'< \gamma$, and since we let the chemical potential be in the range $-2\gamma \leq \mu \leq 2\gamma$, the dominant energy scale in the Hamiltonian $H_0$ is $\gamma$. Therefore, if $\gamma \ll \hbar \Omega$, the condition (\ref{eq:condition}) is satisfied and the Floquet theory may be applied.

In the general case, the effective Hamiltonian is given by \cite{Hemmerich10}
\begin{equation}
H_\textrm{eff} = \bigg \langle \sum_{n=0}^{\infty} \frac{i^n}{n!} \big[ \hat{F}(t), H_0 \big]_n \bigg \rangle_T ,
\label{eq:app-H-eff}
\end{equation}
where for the shaken honeycomb lattice, we choose
\begin{equation}
\hat{F}(t) = \frac{m \Omega^2}{\hbar\Omega} \sin(\Omega t) \left( \sum_{\textbf{r} \in A} \textbf{r}\cdot\boldsymbol{\rho} \, a^\dagger_\textbf{r} a_\textbf{r} + \sum_{\textbf{r} \in B} \textbf{r}\cdot\boldsymbol{\rho} \, b^\dagger_\textbf{r} b_\textbf{r} \right).
\end{equation}
Using the (nonvanishing) commutation relations
\begin{align}
[a^\dagger_{\textbf{r}'}\,a_{\textbf{r}'}, a^\dagger_{\textbf{r}} \, b_{\textbf{r}+\textbf{d}_j} ] =& a^\dagger_{\textbf{r}'} \, b_{\textbf{r}+\textbf{d}_j} \, \delta_{\textbf{r}', \textbf{r}} \,, \nonumber \\
[a^\dagger_{\textbf{r}'}\,a_{\textbf{r}'}, b^\dagger_{\textbf{r}+\textbf{d}_j} \, a_{\textbf{r}}] =& - b^\dagger_{\textbf{r}+\textbf{d}_j} \, a_{\textbf{r}'} \, \delta_{\textbf{r}', \textbf{r}} , \nonumber
\\
[a^\dagger_{\textbf{r}'}\,a_{\textbf{r}'}, a^\dagger_{\textbf{r}} \, a_{\textbf{r}+\textbf{d}_i -\textbf{d}_j}] =& a^\dagger_{\textbf{r}'} \, a_{\textbf{r}+\textbf{d}_i -\textbf{d}_j} \, \delta_{\textbf{r}',\textbf{r}} \nonumber \\
&- a^\dagger_\textbf{r} \, a_{\textbf{r}'} \, \delta_{\textbf{r}',\textbf{r}+\textbf{d}_i -\textbf{d}_j},
\label{eq:commutators}
\end{align}
which are valid for both fermionic and bosonic creation and annihilation operators,
and equivalent ones for the creation and annihilation operators on the $B$ sublattice, the multiple commutator in Eq.\,(\ref{eq:app-H-eff}) becomes
\begin{align}
\big[ \hat{F}(t), H_0 \big]_n &= \left[\frac{m \Omega^2}{\hbar \Omega} \sin(\Omega t) \right]^n \nonumber \\
\bigg\{ &- \gamma \sum_{j=1}^3 \sum_{\textbf{r} \in A} (\textbf{d}_j \cdot \boldsymbol{\rho})^n \left[ (-1)^n a^\dagger_\textbf{r} b_{\textbf{r} + \textbf{d}_j} + b^\dagger_{\textbf{r} + \textbf{d}_j} a_\textbf{r} \right] \nonumber \\
&- \gamma' \sum^3_{i=1} \sum^3_{j=1,j\neq i} \bigg( \sum_{\textbf{r} \in A} [(\textbf{d}_j - \textbf{d}_i) \cdot \boldsymbol{\rho}]^n \, a^\dagger_\textbf{r} \, a_{\textbf{r} + \textbf{d}_i - \textbf{d}_j} \nonumber \\
&+ \sum_{\textbf{r} \in B} [(\textbf{d}_j - \textbf{d}_i) \cdot \boldsymbol{\rho}]^n \, b^\dagger_\textbf{r} b_{\textbf{r} + \textbf{d}_i - \textbf{d}_j} \bigg) \bigg\} \nonumber \\
&- \mu \bigg( \sum_{\textbf{r} \in A} \, a^\dagger_\textbf{r} a_\textbf{r} + \sum_{\textbf{r} \in B} \, b^\dagger_\textbf{r} b_\textbf{r} \bigg) .
\end{align}
Finally, after performing the time average and evaluating the sum over $n$, the effective Hamiltonian (\ref{Hamiltonian-eff}) is obtained.

\end{document}